\documentclass[11pt,article]{elsarticle}

\usepackage{lineno,hyperref}

\usepackage[lmargin=.75in,rmargin=.75in,tmargin=1.in,bmargin=1in]{geometry}
\usepackage{amsmath}
\usepackage{amsthm}
\usepackage{amsfonts}
\usepackage{bbm}
\usepackage{amssymb}
\usepackage{graphicx}
\usepackage{multicol}
\usepackage{wrapfig}
\usepackage{setspace}
\usepackage{epstopdf}
\usepackage{graphicx}
\usepackage{caption}
\usepackage{subcaption}
\usepackage{algorithm}
\usepackage[noend]{algpseudocode}
\usepackage[usenames, dvipsnames]{color}
\newtheorem*{remark}{Remark}
\newtheorem{definition}{Definition}
\usepackage{tabulary}
\usepackage{booktabs}

\usepackage{color}

\journal{Arxiv}

\begin{document}

\begin{frontmatter}

\title{Spatial Poisson processes for fatigue crack initiation}

\author[ICES]{Ivo Babu\v {s}ka}
\ead{babuska@ices.utexas.edu}
\author[KAUST]{Zaid Sawlan \corref{corrauthor}}
\cortext[corrauthor]{Corresponding author}
\ead{zaid.sawlan@kaust.edu.sa}
\author[KAUST,Ur]{Marco Scavino}
\ead{mscavino@iesta.edu.uy}
\author[St]{Barna Szab\'o}
\ead{szabo@wustl.edu}
\author[KAUST]{Ra\'ul Tempone}
\ead{raul.tempone@kaust.edu.sa}

\address[ICES]{The University of Texas at Austin, ICES, Austin, USA}
\address[KAUST]{King Abdullah University of Science and Technology (KAUST), Computer, Electrical and Mathematical Sciences \& Engineering Division (CEMSE), Thuwal 23955-6900, Saudi Arabia}
\address[St]{Washington University in St. Louis, St. Louis, USA}
\address[Ur]{Universidad de la Rep\'ublica, Instituto de Estad\'{\i}stica (IESTA), Montevideo, Uruguay}

\begin{abstract}
In this work we propose a stochastic model for estimating the occurrence of crack initiations on the surface of metallic specimens in fatigue problems that can be applied to a general class of geometries. The stochastic model is based on spatial Poisson processes with intensity function that combines stress-life (S-N) curves with averaged effective stress, $\sigma_{\mathrm{eff}}^{\Delta}(\bold{x})$, which is computed after solving numerically the linear elasticity equations on the specimen domains using finite element methods. Here, $\Delta$ is a parameter that characterizes the size of the neighbors covering the domain boundary. The averaged effective stress, parameterized by $\Delta$, maps the stress tensor to a scalar field upon the specimen domain. Data from fatigue experiments on notched and unnotched sheet specimens of 75S-T6 aluminum alloys are used to calibrate the model parameters for the individual data sets and their combination. Bayesian and classical approaches are applied to estimate the survival-probability function for any specimen tested under a prescribed fatigue experimental setup. Our proposed model can predict the initiation of cracks in specimens made from the same material with new geometries. 
\end{abstract}


\end{frontmatter}



\begin{table}[htbp]\caption{List of notation}
\begin{center}
\begin{tabular}{r p{15cm} }
\toprule
$S-N$ model  & stress-number of cycles model \\
$N$   & 	random number of stress cycles when first crack initiates \\
$S_{max}$     &          maximum stress (in unnotched specimens), nominal maximum stress (in notched specimens) \\
$S_{mean}$ 	& 		nominal mean stress \\
$R$     &          cycle/stress ratio \\
$S_{eq}^{(q)}$ 	& 		equivalent stress with fitting parameter $q$ \\
$W_{max}$ and $W_{min}$  & 		maximum and minimum width of the specimen \\
$T$     &          load force \\
$D$ 	& 		physical domain of the specimen  \\
$\partial D$  &  boundary of the physical domain of the specimen \\
$\bold{u} $ 	& 		displacement field  \\
$E$ 	& 		modulus of elasticity  \\
$\nu $ 	& 		Poisson's ratio  \\
$G$  & 		shear modulus \\
$\sigma_x$ and $\sigma_y$  	& normal stresses in the $x$-axis and $y$-axis  \\
$\tau_{xy}$		& shear stress \\
$\sigma^{1}(\bold{y})$   &  stress tensor with a unit load force at $\bold{y} \in D$   \\
$\sigma(\bold{y})$ 		&     stress tensor at $\bold{y} \in D$ \\
$\sigma_{\mathrm{eff}}(\bold{y})$	& 	effective stress at $\bold{y} \in D$ \\
$\sigma^{\Delta}_{\mathrm{eff}}(\bold{x})$	& 	averaged effective stress over a cube (square) $B(\bold{x},\Delta)$ of length $\Delta$ centered at $\bold{x}$ \\
$\sigma^{\Delta *}_{\mathrm{eff}}$ 	& 	maximum averaged effective stress over the boundary of the physical domain $D$ of the specimen\\
$L(\theta,\Delta; \bold{n})$	& 	likelihood function for the parameter vector $\theta$ of the S-N model and the parameter $\Delta$ given the $m$ observations $\bold{n} = (n_1, \ldots, n_m)$  \\
$g(t; \mu, \sigma)$ 	& 	probability density function of a normal distribution with mean $\mu$ and standard deviation $\sigma$  \\
$\Phi$ 	& 	cumulative distribution function of the standard normal distribution \\
AIC	& 	Akaike information criterion \\
$f_{SN}(n;s,\theta)$ &  failure density function for a given S-N model \\
$F_{SN}(n;s,\theta)$  & cumulative distribution function for a given S-N model \\ 
$h_{SN}(n;s,\theta)$ &   failure/hazard rate function for a given S-N model \\
$ \lambda(\bold{x},n)$  & intensity function of a spatial Poisson process at the location $\bold{x}$ and number of cycles $n$ \\ 
$\eta$  & failure rate function \\
$\lambda _B(n)$  &  rate function of the Poisson process for the number of cracks over a surface region $B$ \\  
$M_B(n)$  & random number of crack initiations in the region $B$ after performing $n$ stress cycles \\
$\Lambda_B(n)$  & cumulative rate function  \\
$N_{\partial D}$  &  random number of stress cycles when the first crack initiates on $\partial D$ \\
$\rho^{\partial D}$  & density function of the random variable $N_{\partial D}$ \\
$\gamma$   &  size of the highly stressed volume of the specimen \\
$\sigma^{1}_{\mathrm{eff}}(\bold{x})$   &  effective stress for a unit load force \\ 
$\mathcal{A}_{\beta}$  & boundary region of the specimen where the effective stress for a unit load force is greater than $\beta$ \\ 
$\gamma(\beta)$   &  highly stressed volume/area \\
$\ell(\theta, \beta, \Delta)$   & 	log-likelihood function \\
\bottomrule
\end{tabular}
\end{center}
\label{tab:TableOfNotationForMyResearch}
\end{table}

\section{Introduction}


Predicting fatigue in mechanical components is extremely important for preventing hazardous situations. Fatigue starts with crack initiation, then the small crack propagates to become a complete fracture. To model the fatigue of real mechanical objects, the geometry of the objects must be considered in the mathematical formulation to compute the stress or strain field. In this work, we are only concerned with crack initiation and use survival analysis to model the time until the first crack occurs. As such, survival means that no cracks are initiated and failure means the initiation of cracks. We focus only on stress-based approaches that model fatigue crack initiation under high cycle fatigue.





Our goal is to construct a stochastic model that can estimate the survival probability of any mechanical component given data of fatigue experiments on specimens made of the same material. Statistical S-N models are usually used for survival prediction in uniaxial-fatigue experiments with cyclic loadings \cite{fatigue1, leonetti2017fitting}. There are also several generalizations of S-N models for fatigue of notched specimens (see \cite{szabo} and the references therein). In these generalizations, the stress tensor field is computed around the notch and mapped into a stress predictor characterized by some parameters. Instead, we consider an effective stress field over the full domain and assume the existence of a material parameter $\Delta$ that define a local neighbor to average the effective stress locally following the theory of critical distances (TCD). There are many versions of the TCD involving additional material parameters \cite{susmel2007novel,taylor,taylor2010}. Considering the limited amount of available data and the goal of this study, the most simple TCD using only the averages was used and calibrated.

In this work, we develop a general approach, following the ideas in \cite{banff}, to modeling crack initiation on the surface of metallic specimens based on simple assumptions. Our model transforms the stress tensor field into a spatial stress function named the averaged effective stress function, $\sigma_{\mathrm{eff}}^{\Delta}(\bold{x})$. We combine $\sigma_{\mathrm{eff}}^{\Delta}(\bold{x})$ with a chosen S-N model to build a rate function for the spatial Poisson process. The rate function is scaled by the highly stressed volume that is parameterized by a threshold parameter, $\beta$. The choices of $\sigma_{\mathrm{eff}}^{\Delta}(\bold{x})$, the S-N model and the parameterization of the highly stressed volume are subject to user preference. The resulting model is independent of the shape of the specimen and can fit fatigue data from both notched and unnotched specimens. The Poisson process has been used similarly with a local strain field for low cycle fatigue with unnotched specimens in \cite{schmitz2013probabilistic} and recently with notched specimens in \cite{made2018combined}. We underline that references \cite{schmitz2013probabilistic} and \cite{made2018combined} are concerned with low cycle fatigue whereas the present paper is concerned with high cycle fatigue.

Following our previous work \cite{fatigue1}, we first calibrate the model parameters by means of the maximum likelihood (ML) method, and we consider Bayesian analysis to provide a better understanding of the results obtained by classical methods. The Bayesian analysis has become a primary part of studying fatigue life prediction. For example, Bayesian methods were used to analyze a new S-N model with fatigue tests on welded cover-plate steel-beam specimens \cite{leonetti2017fitting} and low-cycle fatigue models for turbine disks \cite{wang2017zone}. In \cite{faghihi2018probabilistic}, the Bayesian framework was used to assess the uncertainty of a continuum damage model. A hierarchical Bayesian approach was also used in \cite{liu2017hierarchical}, which allowed general prior models to be considered.

The rest of the paper is organized as follows. In Section \ref{data}, we analyze available data from fatigue experiments to determine their utility in calibrating crack-initiation models. Section \ref{effstress} includes a description of the linear elasticity equations used to compute the stress tensor. Then, we consider a map that transforms the stress tensor into a real-valued spatial function (the averaged effective stress). In Section \ref{calb}, we recall a simple S-N model and calibrate it to the data introduced in Section \ref{data} by assuming that the first crack will initiate at the spatial position with the highest effective stress. We introduce our novel spatial Poisson model in Section \ref{Poisson} and derive the exact form of the log-likelihood function. This new model is extensively applied to estimate the survival-probability for each specimen in the data. Finally, we analyze our Poisson model via a Bayesian framework in Section \ref{Bayesian} where the survival functions computed by posterior estimates for the entire set of data (hereafter referred to as pooled posterior estimates) are compared with the reference ML estimate of each data set. 

\section{Data sets from experiments on aluminum-sheet specimens}
\label{data}
\begin{figure}[h!]
\centering
\includegraphics[width=1.0\textwidth]{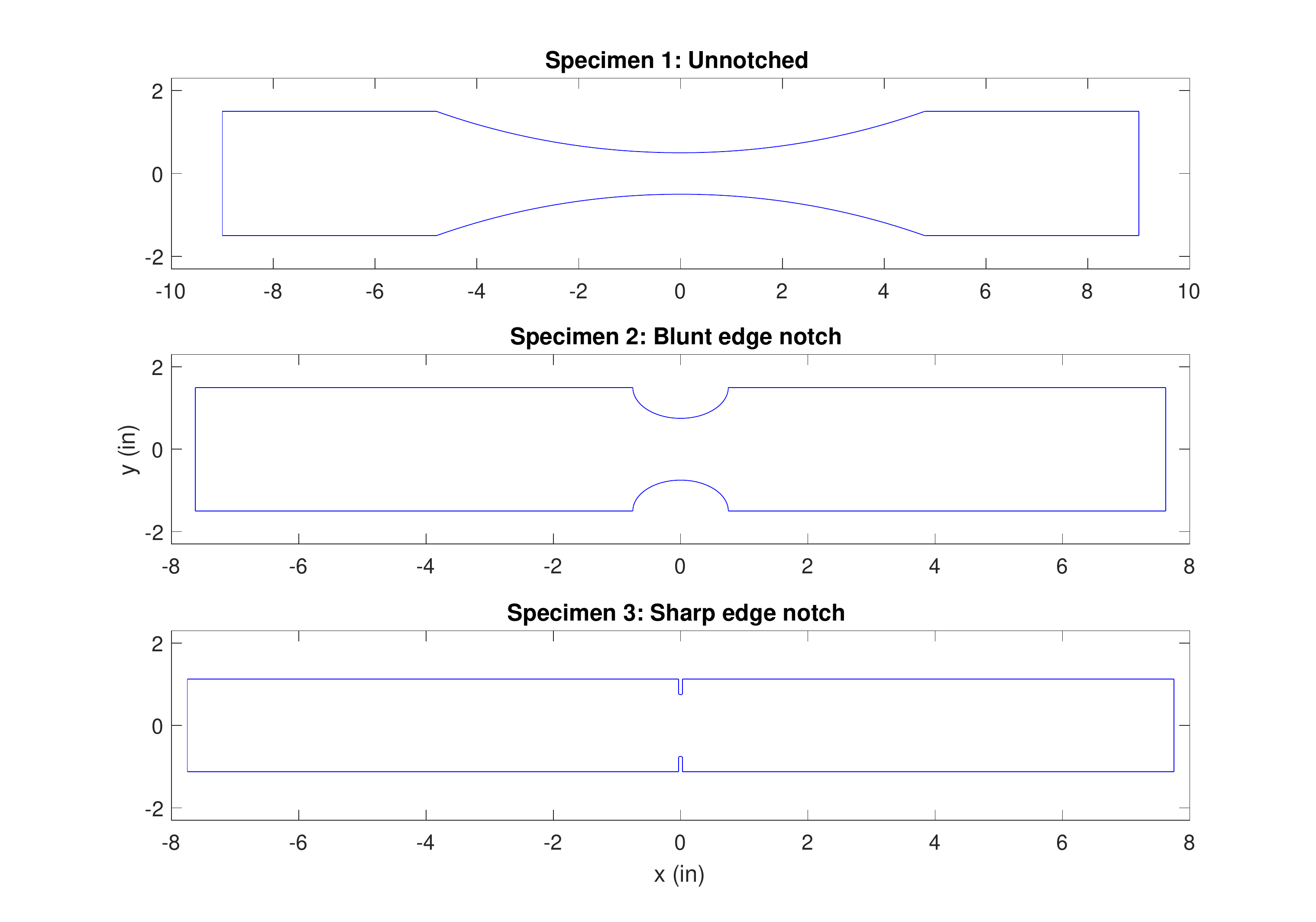}
\caption{Three different types of 75S-T6 aluminum specimens (\cite{TN2324,TN2639,TN2390}).}
\label{specimens}
\end{figure}

We extract three data sets from the National Advisory Committee for Aeronautics (NACA) technical notes 2324 \cite{TN2324}, 2639 \cite{TN2639} and 2390 \cite{TN2390}, which correspond to fatigue experiments conducted on three sheets of 75S-T6 aluminum (Figure \ref{specimens}). Specimen 1 is an unnotched, dogbone-shaped specimen, while specimens 2 and 3 are notched with blunt and sharp notches, respectively.

The first data set consists of fatigue experiments applied to specimen 1 \cite[table 3, pp.22--24]{TN2324}. In each experiment, the following data are recorded:
\begin{itemize}
\item The maximum stress, $S_{max}$, measured in ksi units.
\item The cycle ratio, $R$, defined as the minimum to maximum stress ratio.
\item The number of load/stress cycles at which fatigue failure occurred.
\item A binary variable (0/1) denoting whether or not the experiment was stopped prior to failure (run-out).
\end{itemize}

The other data sets contain fatigue experiments applied to the notched specimens (see \cite[table 3, p.9]{TN2639} and \cite[table 3, p.8]{TN2390}). We consider only experiments with high cycle fatigue; that is, $N > 8800$. In these data sets, the following data are recorded:
\begin{itemize}
\item The nominal maximum stress $S_{max}$, defined as the maximum applied force divided by the smallest cross-sectional area of the test specimen.
\item the nominal mean stress, $S_{mean}$, the stress ratio is then given by $R = 2\frac{S_{mean}}{S_{max}} -1$.
\item The number of load/stress cycles at which fatigue failure occurred.
\item A binary variable (0/1) denoting whether or not the experiment was stopped prior to failure (run-out).
\end{itemize}

\begin{remark}
The fatigue data obtained for particular cycle ratios must be generalized to arbitrary cycle ratios. For this purpose, the equivalent stress is defined as 
\begin{equation} \label{F2Seq}
S_{eq}^{(q)} = S_{max}\,(1-R)^{q},
\end{equation} 
where $q$ is a fitting parameter \cite{fatigue1, faa, walker}. The load force, or normal traction, is then given by $T = \frac{W_{min}}{W_{max}}S_{eq}^{(q)}$, where $W_{max}$ is the maximum width of the specimen and $W_{min}$ is the minimum width.
\end{remark}

\begin{table}[h!]
\begin{center}
\caption{Fatigue data for 75S-T6 aluminum alloys.} 
\begin{tabular}{|c|c|c|c|c|}
\hline
Data set & Specimen & Source & Radius (in) & Number of experiments \\
\hline   
1 & 1 (Unnotched) & NACA TN 2324 \cite{TN2324} & 12 & 85 (73) \\
\hline    
2 & 2 (Blunt edge notch) & NACA TN 2639 \cite{TN2639} & 0.76 & 31 (29) \\
\hline
3 & 3 (Sharp edge notch) & NACA TN 2390 \cite{TN2390} & 0.03125 & 28 (22) \\
\hline
\end{tabular}
\label{Data}
\end{center}
\end{table} 

Table \ref{Data} summarizes the properties of each specimen shown in Figure \ref{specimens}. The total number of experiments is also given with the number of observed failures provided in parentheses. The thickness of all three specimens is $0.09$ in, which is relatively small. Therefore, we can reduce the dimension of the problem to two. However, we note that for specimen 3, the radius of the notch is smaller than the specimen thickness; therefore, the three-dimensional (3D) model might not be sufficiently approximated by a two-dimensional (2D) model.  

\begin{remark}
In the aforementioned experimental data, we assume that crack propagation is instantaneous as crack initiation occurs; that is, crack initiation is equivalent to fatigue failure. We validate this assumption by Paris’ law for some experiments. The number of cycles spent on crack propagation is negligible compared with the number of cycles until crack initiation. The life of the specimen is thus defined as the number of stress cycles until the first crack initiates. This assumption allows us to calibrate fatigue crack-initiation models to the available data. In general, we would need the number of stress cycles when the first crack appears.
\end{remark}

\section{Numerical computation of $\sigma^{\Delta}_{\mathrm{eff}}(\bold{x})$}
\label{effstress}
\subsection{Linear elasticity}
\begin{figure}[h!]
\centering
\includegraphics[width=0.8\textwidth]{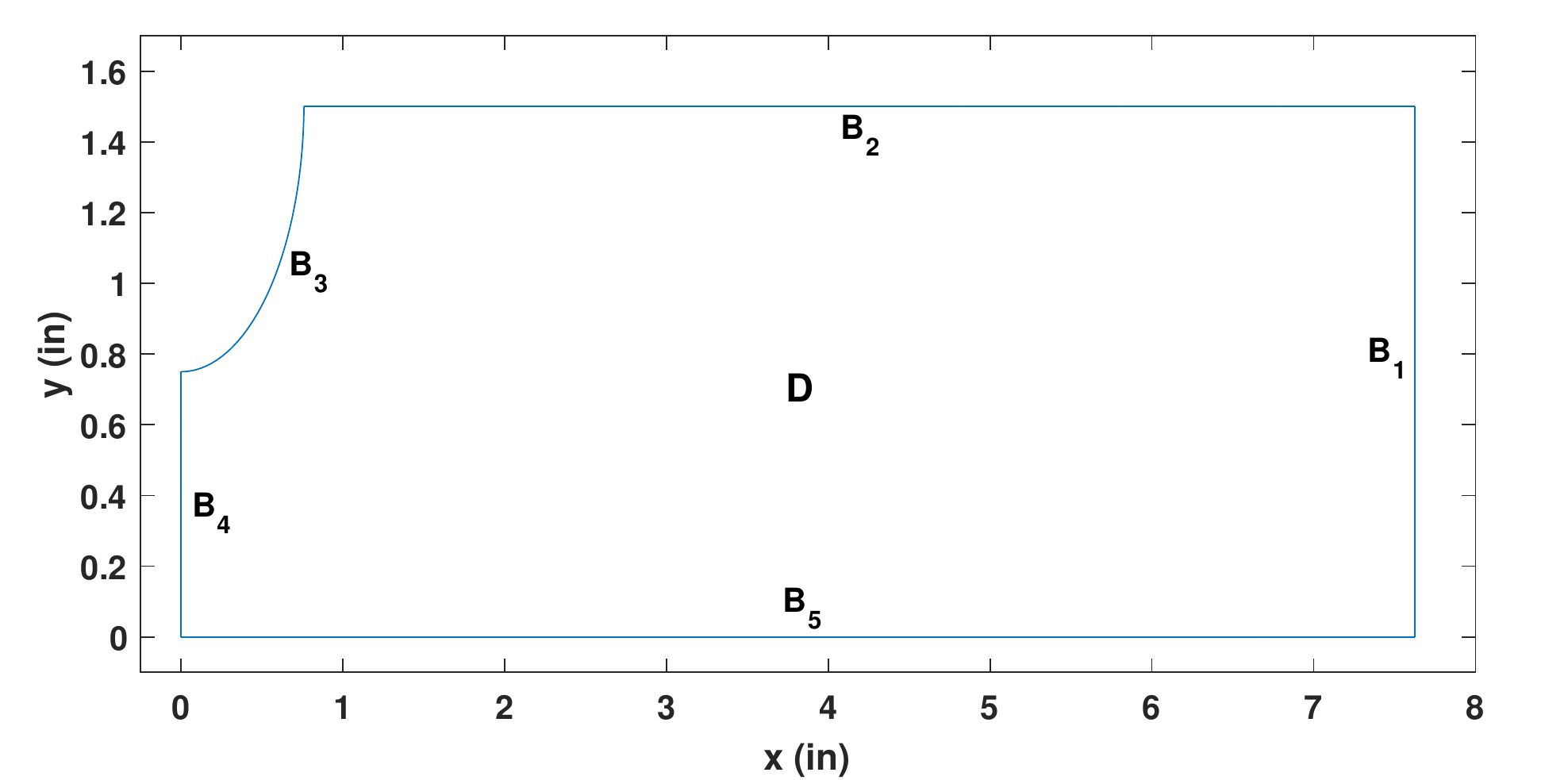}
\caption{Physical domain of quarter of specimen 2.}
\label{domain}
\end{figure}

The stress field in the specimens is defined by the linear elasticity theory. The mathematical model of linear elasticity is based on strain-displacement equations, stress-strain equations, and equilibrium equations \cite{FEM}. We let $D$ be the domain shown in Figure \ref{domain}. Then, the displacement field $\bold{u} = [\begin{array}{cc} u_x(x,y) & u_y(x,y) \end{array}]'$ satisfies the equations of two-dimensional elasticity (plane stress) \cite{timoshenko1970theory}:
\begin{eqnarray*}
\frac{E}{2(1-\nu)} \frac{\partial}{\partial x} \left( \frac{\partial u_x}{\partial x} + \frac{\partial u_{y}}{\partial y} \right) + G \left( \frac{\partial^2 u_x}{\partial x^2} + \frac{\partial^2 u_x}{\partial y^2} \right) = 0 \\
\frac{E}{2(1-\nu)} \frac{\partial}{\partial y} \left( \frac{\partial u_x}{\partial x} + \frac{\partial u_{y}}{\partial y} \right) + G \left( \frac{\partial^2 u_y}{\partial x^2} + \frac{\partial^2 u_y}{\partial y^2} \right) = 0
\end{eqnarray*}
where $E>0$ is the modulus of elasticity, $\nu$ is Poisson's ratio, and $G=\frac{E}{2(1+\nu)}$ is the shear modulus. Equivalently, the two-dimensional elasticity equations can be written in terms of the stress as follows:
\begin{equation}
\label{elasEq}
\begin{cases}
\frac{\partial \sigma_x}{\partial x} + \frac{\partial \tau_{xy}}{\partial y} = 0 \\
\frac{\partial \tau_{xy}}{\partial x} + \frac{\partial \sigma_y}{\partial y} = 0 
\end{cases}
\end{equation}
where the normal stresses in the x-axis and y-axis, $\sigma_x$ and $\sigma_y$, respectively, and the shear stress, $\tau_{xy}$, are given by:
\begin{eqnarray*}
\sigma_x &=& \frac{E}{1-\nu^2} \left( \frac{\partial u_x}{\partial x} + \nu \frac{\partial u_y}{\partial y} \right) \\
\sigma_y &=& \frac{E}{1-\nu^2} \left(\nu \frac{\partial u_x}{\partial x} + \frac{\partial u_y}{\partial y} \right) \\
\tau_{xy} &=& G \left( \frac{\partial u_x}{\partial y} + \frac{\partial u_y}{\partial x} \right)
\end{eqnarray*}
The boundary conditions for the domain $D$ shown in Figure \ref{domain} are $\sigma_x = 1, \tau_{xy} = 0$ on boundary segment $B_1$, free boundary condition are prescribed on $B_2$ and $B_3$, and symmetry boundary conditions are prescribed on $B_4$ and $B_5$.

\begin{figure}[h!]
\centering
\includegraphics[width=18cm]{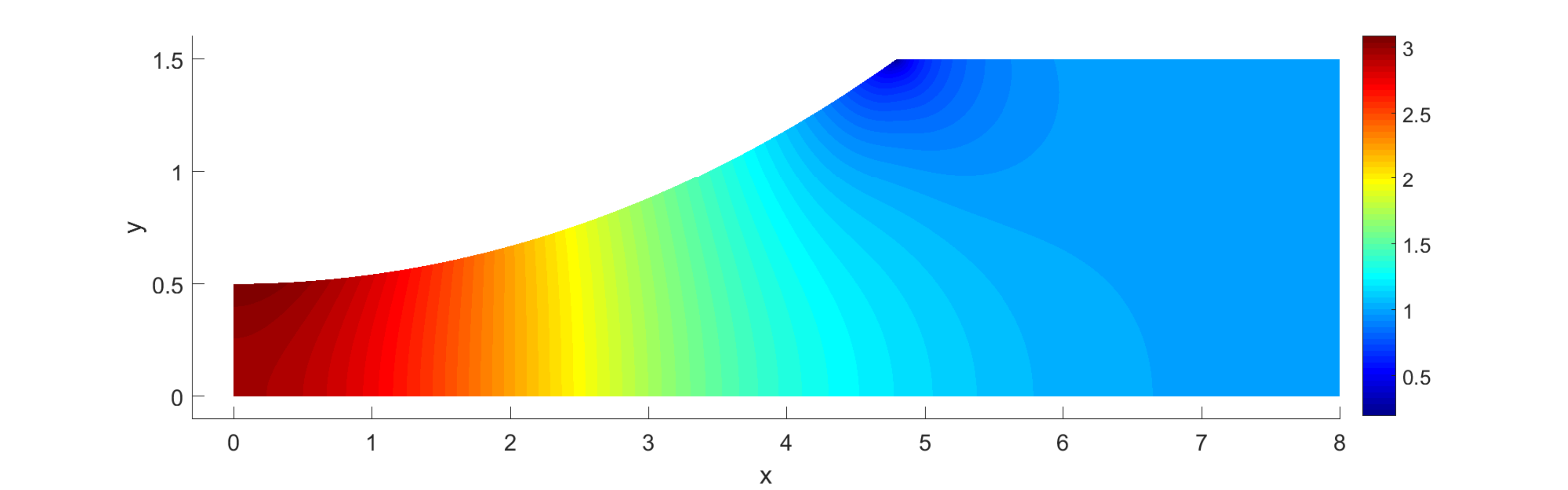}
\caption{Distribution of $\sigma_{x}$ for the unnotched specimen used in NACA TN 2324 (specimen 1).}
\label{Stress1}
\end{figure}

\begin{figure}[h!]
\centering
\includegraphics[width=18cm]{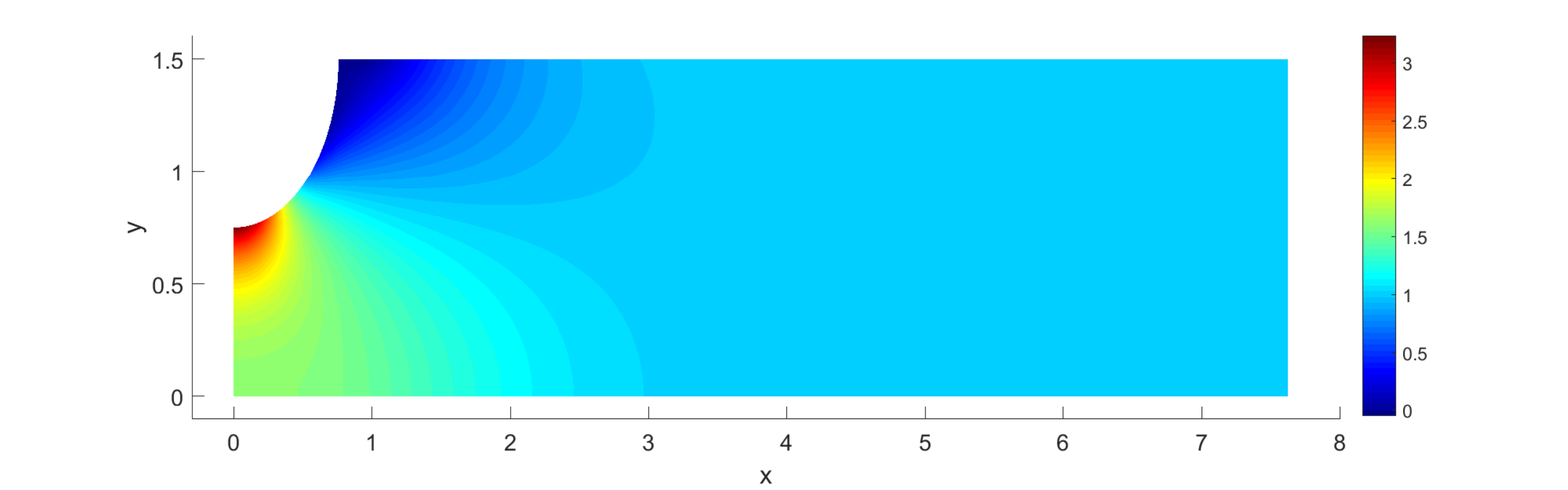}
\caption{Distribution of $\sigma_{x}$ for the notched specimen used in NACA TN 2639 (specimen 2).}
\label{Stress2}
\end{figure}

\begin{figure}[h!]
\centering
\includegraphics[width=16cm]{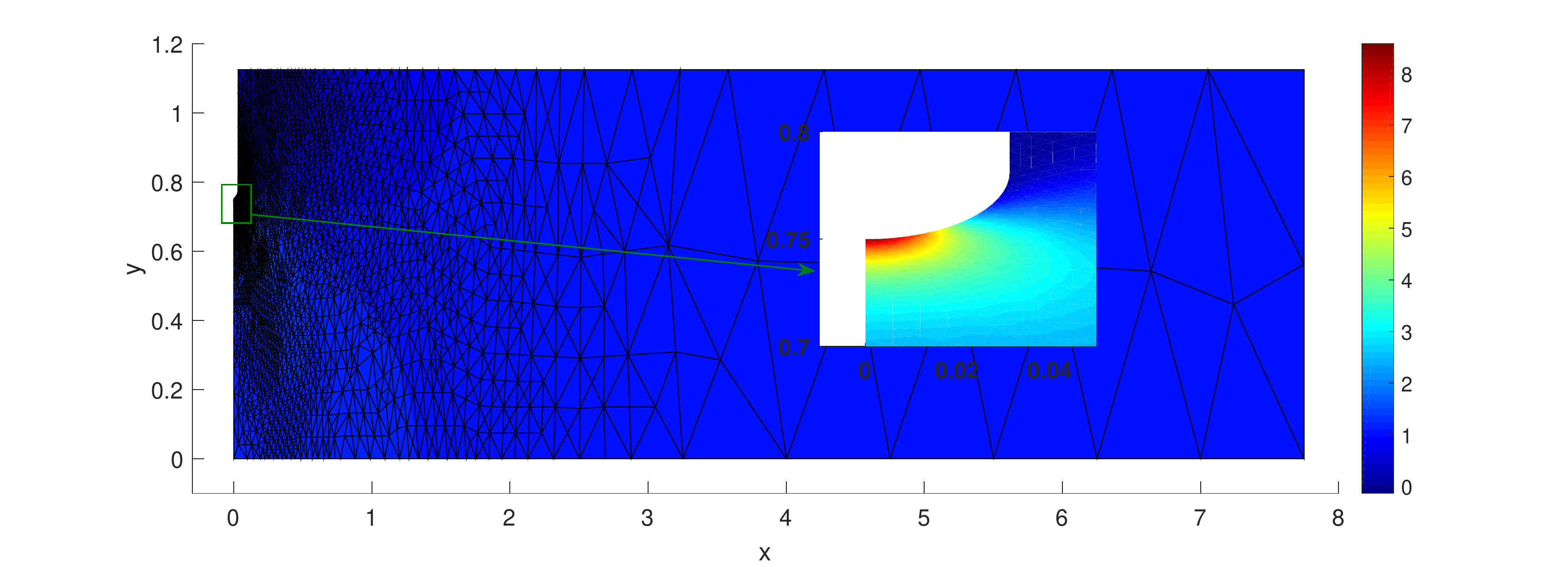}
\caption{Distribution of $\sigma_{x}$ in the plane of symmetry of the notched specimen used in NACA TN 2390 (specimen 3).}
\label{Stress3}
\end{figure}

\begin{figure}[h!]
\centering
\includegraphics[width=16cm]{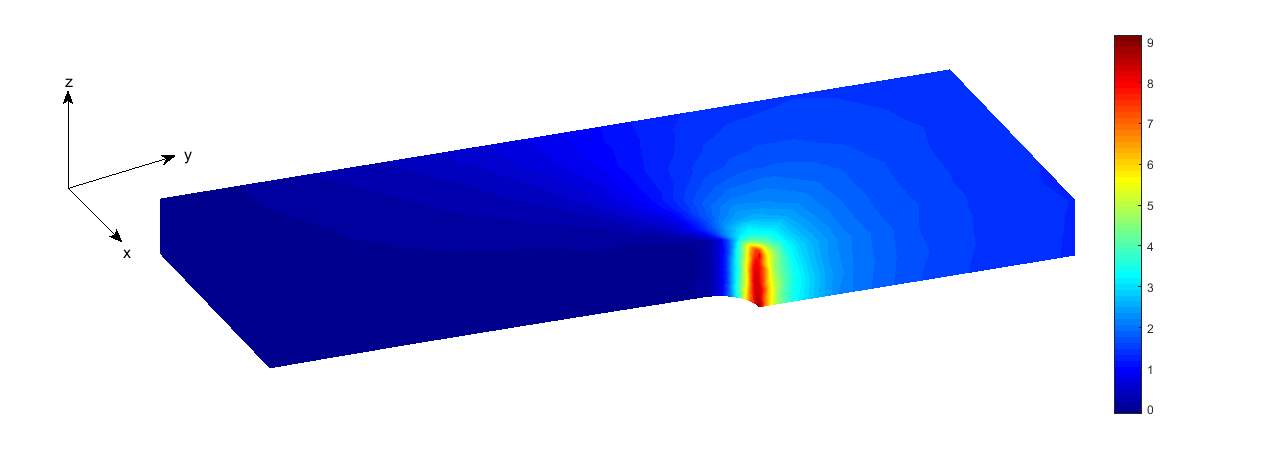}
\caption{Three-dimensional $\sigma_{x}$ of specimen 3; the bottom plane is the plane of symmetry displayed in Figure \ref{Stress3}.}
\label{Stress4}
\end{figure}

The numerical solution for the system \eqref{elasEq} was obtained by the Finite Element Method \cite{FEM}. Figures \ref{Stress1} and \ref{Stress2} show the computed values of $\sigma_{x}$ over the right upper quarter of specimens 1 and 2, respectively. The mesh of the right upper quarter of specimen 3 is shown in Figure \ref{Stress3}. The adaptive PDE solver ADAPTMESH in MATLAB was used. For specimen 3, we also computed three-dimensional stress field using the MATLAB function SOLVEPDE with unstructured mesh generated by COMSOL. The stress $\sigma_x$ is shown in Figure \ref{Stress4}. We refer the reader to \cite[Chapter 3]{FEM} for the details of the three-dimensional elasticity equations.

\begin{remark}
Due to the linearity of the problem, it is sufficient to compute the stress tensor with a unit load force, denoted by $\sigma^{1}(\bold{y})$, $\bold{y} \in D$, for a given specimen. Then, the stress tensor that corresponds to a specific experiment is given by
\begin{equation}
\sigma(\bold{y}) = T \times \sigma^{1}(\bold{y})  = \frac{W_{min}}{W_{max}}S_{max}(1-R)^{q} \times \sigma^{1}(\bold{y}),
\label{stressi}
\end{equation}
where $S_{max}(1-R)^{q}$ is the equivalent stress that characterizes the stress cycle of the experiment as defined in \eqref{F2Seq}. Thus, we need to run a MATLAB PDE solver only once for each specimen.
\end{remark}

In the rest of this section, we will denote the stress tensor at $\bold{y} \in D$ by $\sigma(\bold{y})$, where $\sigma(\bold{y})$ depends on the stress ratio $R$ and the parameter $q$ through equation \eqref{stressi}.

\subsection{Averaged effective stress $\sigma^{\Delta}_{\mathrm{eff}}(\bold{x})$}

There are several proposals for the definition of effective stress \cite{szabo, banff}. In this work, we define the effective stress by the maximum principal stress. The two-dimensional maximum principal stress is given by \cite{stressformulas}:
\begin{equation*}
\sigma_{\mathrm{eff}}(x,y) = \frac{1}{2} (\sigma_x + \sigma_y) + \sqrt{(\frac{\sigma_x - \sigma_y}{2})^2 + \tau_{xy}^2}\, , \forall (x,y) \in D.
\end{equation*}

Then, we obtain the averaged effective stress by averaging the effective stress locally:
\begin{equation}
\sigma^{\Delta}_{\mathrm{eff}}(\bold{x}) = \frac{1}{|B(\bold{x},\Delta) \cap D|} \int_{B(\bold{x},\Delta) \cap D} \sigma_{\mathrm{eff}}(\bold{y}) d\bold{y} \,.
\end{equation}
where $B(\bold{x},\Delta)$ is a cube (square) of length $\Delta$ centered at $\bold{x}$ \cite{banff} in the 3D (2D) model. In general, there are different ways to define $B(\bold{x},\Delta)$, as in the theory of critical distances \cite{taylor}. As $\Delta$ converges to zero, the averaged effective stress converges to the effective stress.

\section{Preliminary calibration}
\label{calb}
In our initial attempt to model fatigue crack initiation, we assume that the crack formation is determined by $\sigma^{\Delta *}_{\mathrm{eff}} = \max_{\bold{x} \in \partial D} \sigma^{\Delta}_{\mathrm{eff}}(\bold{x})$. This assumption reduces the stress field to a scalar value that can be used with S-N models. We consider the fatigue-limit model (Model Ia) from \cite{fatigue1}. The fatigue life, $N$, is modeled by means of a lognormal distribution with mean $\mu(\sigma^{\Delta *}_{\mathrm{eff}})$ and standard deviation $\tau$, where
\begin{equation*}
\mu(\sigma^{\Delta *}_{\mathrm{eff}}) = 
\begin{cases}
A_1 + A_2 \, \log_{10}( \sigma^{\Delta *}_{\mathrm{eff}} - A_3)\,,\:\:\textrm{if}\:\:\sigma^{\Delta *}_{\mathrm{eff}} > A_3 \\
+\infty \,,\:\:\:\:\:\:\textrm{otherwise}
\end{cases}
\end{equation*}
and $A_3$ is the fatigue-limit parameter. When $\mu = +\infty$, we assume cracks will never initiate and the survival probability will be constant equals $1$. The likelihood function for the S-N curve based on $m$ experiments is given by
\begin{equation}
\label{SNIa}
L(\theta,\Delta; \bold{n}) = \prod_{i=1}^{m} \left[ \frac{1}{n_i \log(10)} g(\log_{10}(n_i)\,;\mu( \sigma^{\Delta *}_{\mathrm{eff},i})\,,\tau) \right]^{\delta_i} \, 
\left[ 1- \Phi \left( \frac{\log_{10}(n_i) - \mu( \sigma^{\Delta *}_{\mathrm{eff},i})}{\tau} \right) \right]^{1 - \delta_i}\,,
\end{equation}
where $\theta = (A_1, A_2, A_3, q, \tau)$, $\bold{n} = (n_1, \ldots, n_m)$, $g(t; \mu, \sigma) = \frac{1}{\sqrt{2 \pi} \,\sigma} exp \left\{ - \frac{(t - \mu)^2}{2 \sigma^2} \right\}\,,$ $\Phi$ is the cumulative distribution function of the standard normal distribution, and
\begin{equation*}
\delta_i = \left\{
\begin{array}{rl}
1 & \text{if } n_i \text{ is a failure}\\
0 & \text{if } n_i \text{ is a run-out\,.}
\end{array} \right.
\end{equation*} 

\begin{table}[h!]
\begin{center}
\caption{ML estimates of $\theta$ from Model Ia using\eqref{SNIa} where $\Delta = 0$.}
\begin{tabular}{|c|c|c|c|c|c|c|c|}
\hline
Data set & $A_1$ & $A_2$ & $A_3$ & $q$ & $\tau$  & Maximum log-likelihood & AIC \\
\hline
1 & 7.40 &  -2.01  &  35.91  &  0.5627  &  0.5274 & -950.16 & 1910.32 \\
\hline
2 & 8.24 &  -2.39  &  38.98  &  0.5790  &  0.4216  & -396.55 & 803.10 \\
\hline
3 (2D) & 8.78 &  -2.46  &  46.41  &  0.6470  &  0.5441 &  -303.44 &  616.88 \\
\hline
3 (3D) & 8.86 &  -2.47  &  48.94  &  0.6470  &  0.5439 &  -303.44 & 616.88 \\
\hline
1, 2 \& 3 (2D) & 7.77 &  -1.96  &  35.58  &  0.5780  &  0.7909 &  -1710.68 & 3431.36 \\
\hline
1, 2 \& 3 (3D) & 7.50 &  -1.75  &  36.11  &  0.5755  &  0.8199 &  -1715.12 &  3440.24 \\
\hline
\end{tabular}
\label{mle1anotch}
\end{center}
\end{table}

\begin{figure}[!htb]
\centering
\minipage{0.9\textwidth}
\includegraphics[width=1.0\linewidth]{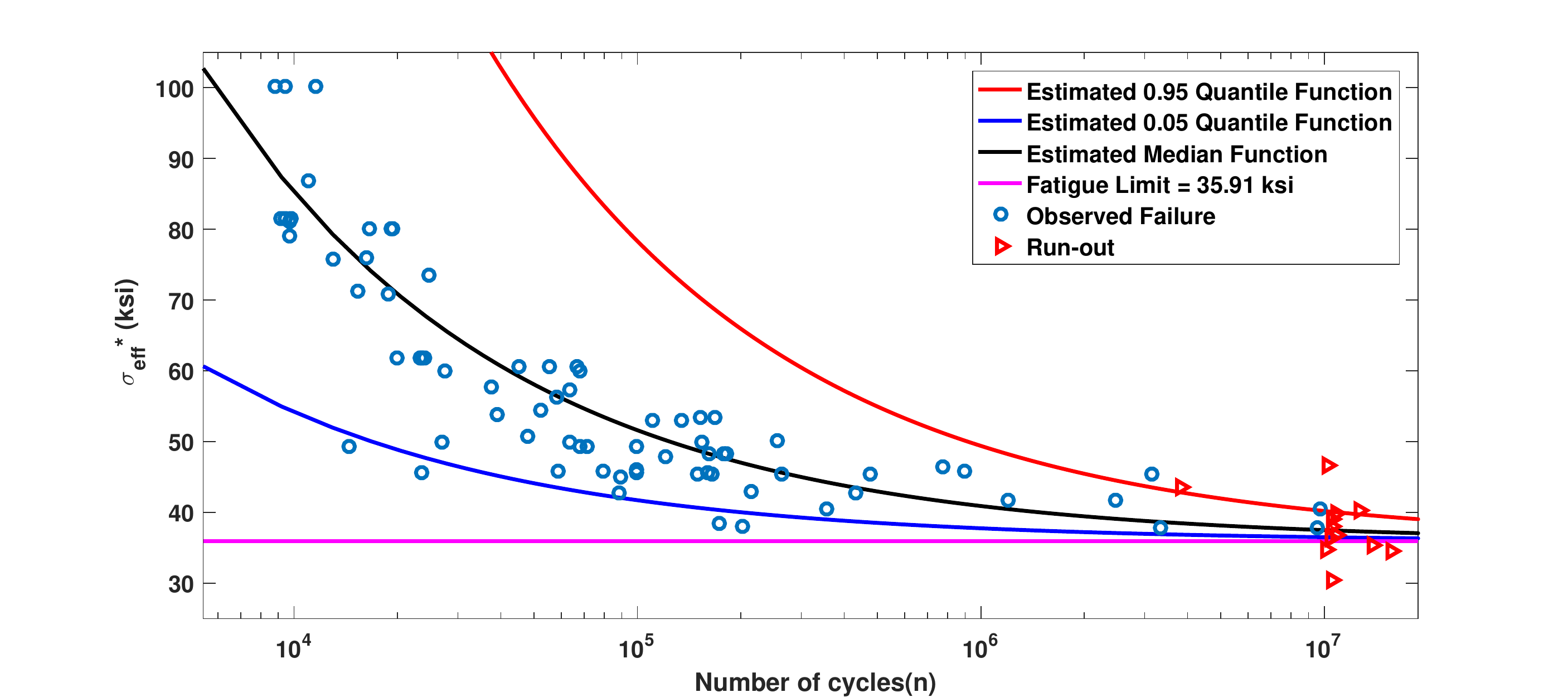}
\endminipage\\
\minipage{0.9\textwidth}
\includegraphics[width=1.0\linewidth]{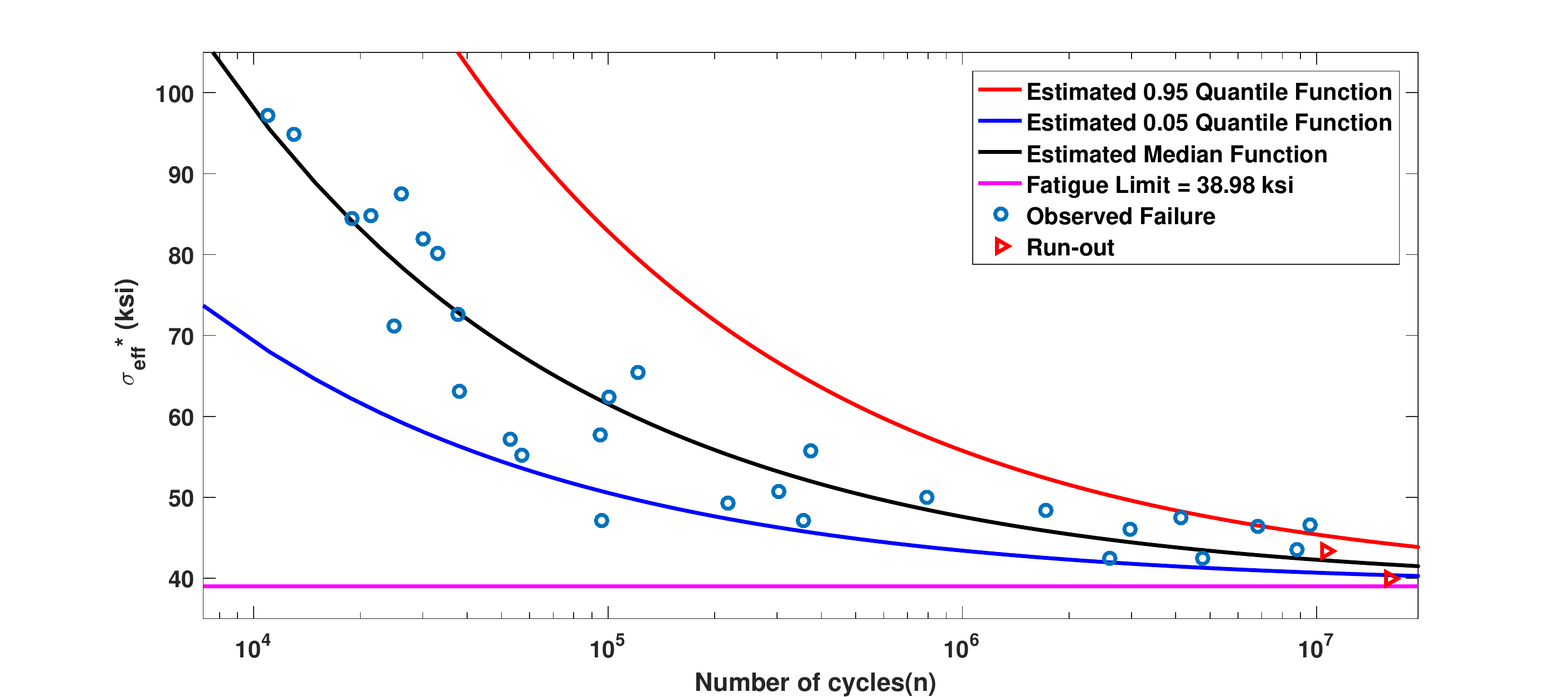}
\endminipage
\caption{Fitting Model Ia to data set 1 (upper panel) and data set 2 (bottom panel) where $\Delta = 0$.}
\label{pc12}
\end{figure}

\begin{figure}[!htb]
\centering
\minipage{0.9\textwidth}
\includegraphics[width=1.0\linewidth]{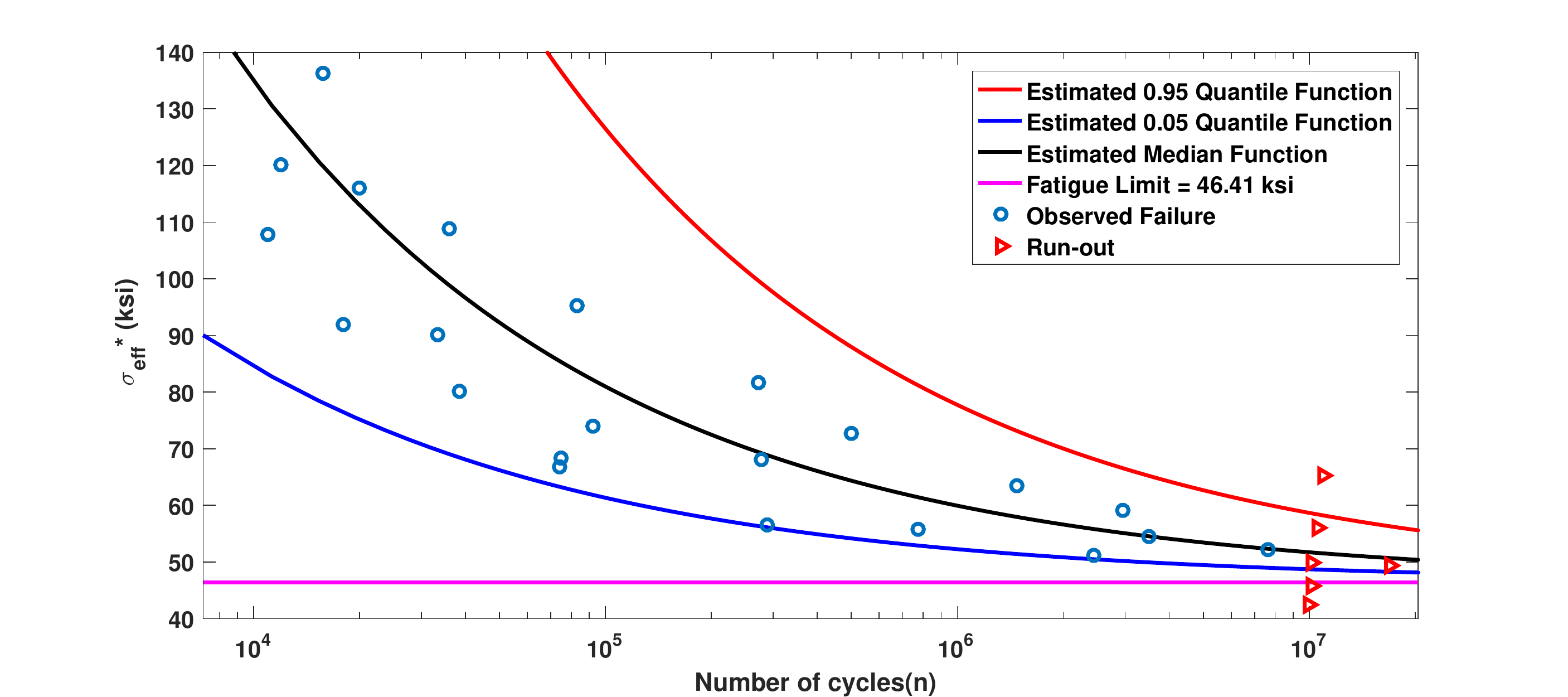}
\endminipage\\
\minipage{0.9\textwidth}
\includegraphics[width=1.0\linewidth]{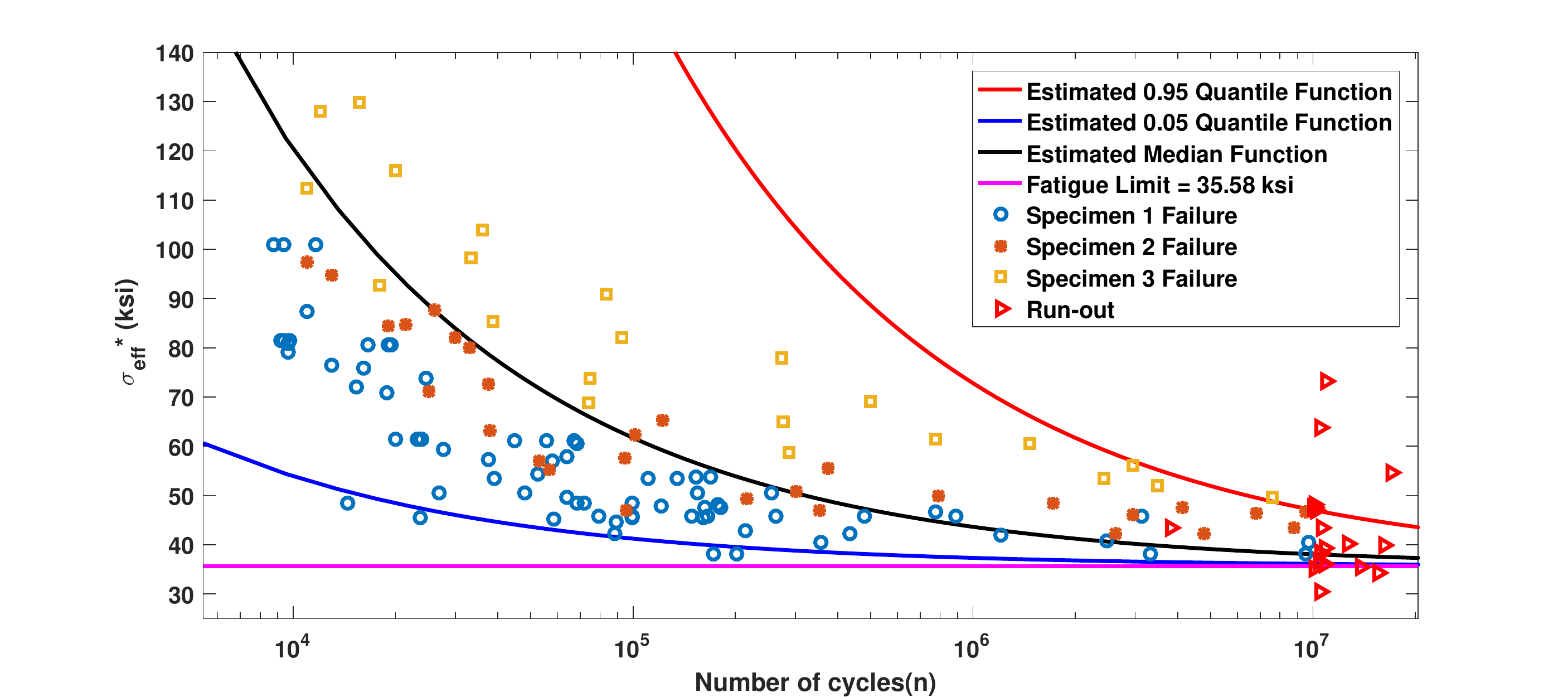}
\endminipage
\caption{Fitting Model Ia to data set 3 (2D) (upper panel) and the combined data set (bottom panel) where $\Delta = 0$.}
\label{pc34}
\end{figure}

Before calibrating $\Delta$, we estimate $\theta = (A_1, A_2, A_3, q, \tau)$ under the restriction that $\Delta = 0$. Table \ref{mle1anotch} shows the corresponding ML estimates of $\theta$ obtained using each data set individually and the pooled ML estimates that are obtained from the combined data set. Figures \ref{pc12} and \ref{pc34} show the corresponding quantile functions with the data used to fit the model. In these figures, the observations are plotted with respect to the maximum effective stress in the y-axis not the applied traction. The notched specimens are subjected to much smaller traction forces but the effective stress will include the notch effect. We notice that the ML estimates of $\theta$ differ considerably with each data set and the variance is increased when fitting the same S-N model to the combined data set. To reduce the variance, more parameters need to be introduced. However, for predicting the life of new specimens, it is preferable to have large variability in the model.

For specimen 3, we consider both 2D and 3D models and show the calibrated parameters for each case in Table \ref{mle1anotch}. When calibrating the specimen 3 data set only, the 2D and 3D models produce the same fit with only a small change in the fatigue-limit parameter. Moreover, the pooled estimates for the combined data sets are very similar, though the 2D model provides a better fit by underestimating the maximum effective stress, which makes it more similar to the maximum effective stress of specimens 1 and 2.
 
\begin{table}[h!]
\begin{center}
\caption{ML estimate of $\theta$ and $\Delta$ from Model Ia using \eqref{SNIa}.}
\begin{tabular}{|c|c|c|c|c|c|c|c|c|}
\hline
Data set & $A_1$ & $A_2$ & $A_3$ & $q$ & $\tau$  & $\Delta$ (in) & Maximum log-likelihood & AIC \\
\hline
1, 2 \& 3 (2D) & 8.16 &  -2.38  &  34.65  &  0.5774  &  0.6006 &  0.025 & -1673.56 & 3359.12 \\
\hline
\end{tabular}
\label{mleDelta}
\end{center}
\end{table}

\begin{figure}[h!]
\centering
\includegraphics[width=16cm]{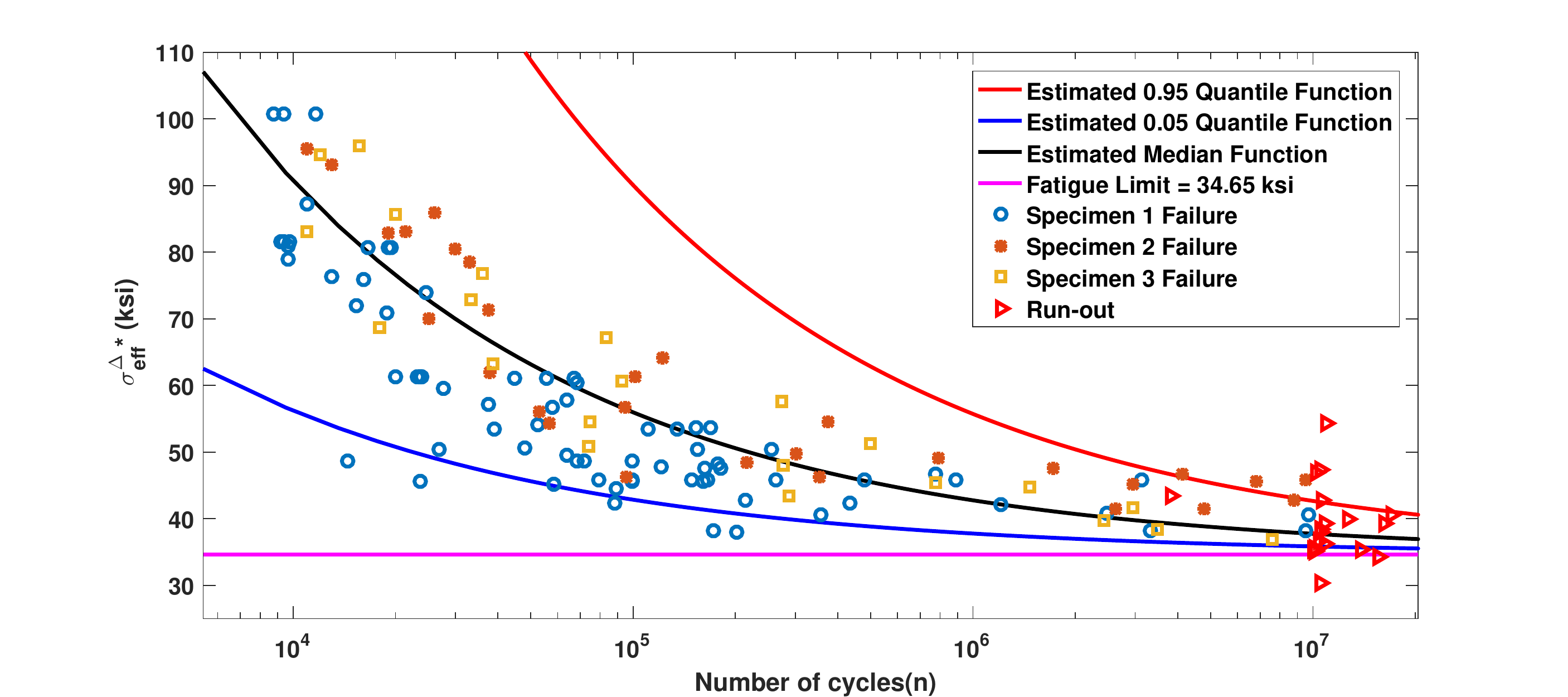}
\caption{Fitting Model Ia to the combined data set where $\Delta > 0$.}
\label{pc5}
\end{figure}

Now, we incorporate the parameter $\Delta$ to average the effective stress locally. By maximizing the likelihood function \eqref{SNIa} using the combined data, we obtain a pooled ML estimate of $\theta$ and $\Delta$ in Table \ref{mleDelta}. We present this fitting in Figure \ref{pc5}. The incorporation of $\Delta$ unifies the range of $\sigma^{\Delta *}_{\mathrm{eff}}$ for the three specimens, and therefore, improves the fitting of the data.

\begin{figure}[h!]
\centering
\includegraphics[width=9cm]{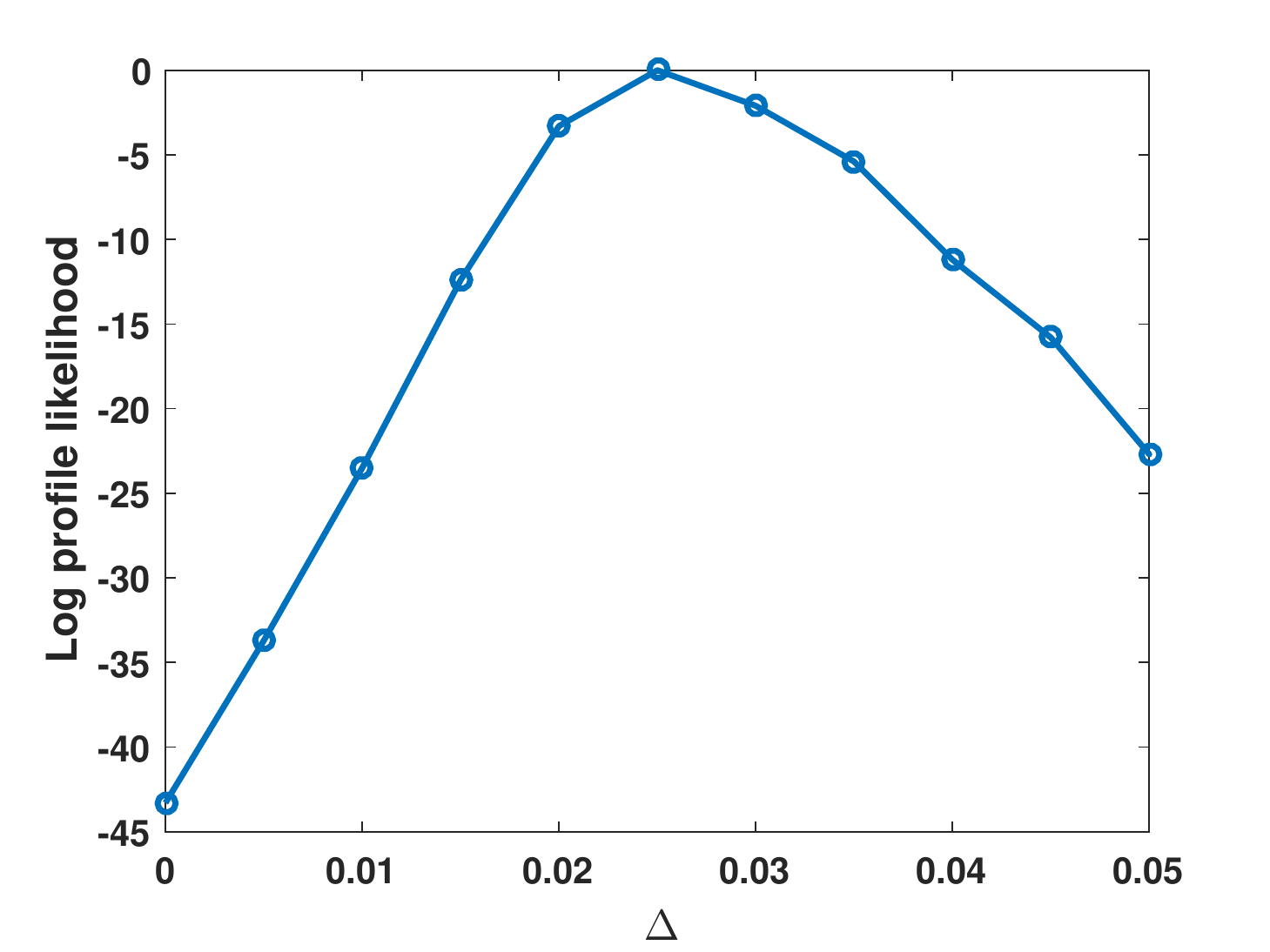}
\caption{The estimated log profile likelihood function for the parameter $\Delta$ derived from \eqref{SNIa} and the combined data sets.}
\label{logprofLike}
\end{figure}

Figure \ref{logprofLike} shows the log of the profile likelihood function for $\Delta$ where the $95\%$ confidence interval is approximately $(0.021, 0.03)$. We compare the fitting of the combined data set when $\Delta = 0$ and $\Delta >0$ by means of the Akaike information criterion (AIC) \cite{aic} in Table \ref{AIC1}. We recall that the AIC is define as $2(p - \log L^*)$ where $p$ is the number of parameters and $L^*$ is the maximum likelihood value. The smaller AIC value corresponds to the better fit. In the rest of the paper, we will use AIC to compare different models with different number of parameters.

\begin{table}[h!]
\begin{center}
\caption{Comparison between two different specifications of Model Ia using \eqref{SNIa}.}
\begin{tabular}{|c|c|c|c|}
\hline
Model Ia given data sets 1, 2 \& 3 (2D)  & $\Delta = 0$ &  $\Delta > 0$ \\
\hline
Maximum log-likelihood &  -1710.68   &  -1673.56 \\
\hline
Akaike information criterion (AIC) &  3431.36  &  3359.12 \\
\hline
\end{tabular}
\label{AIC1}
\end{center}
\end{table}

\begin{remark}
The parameter $\Delta$ smooths the effective stress locally and acts as a scaling parameter for a given non-uniform stress field. Therefore, $\Delta$ must be calibrated using combined data from fatigue experiments on specimens with different geometries. 
\end{remark}

Because it was experimentally observed that the crack was not always at the point where the notch is deepest, in the next section, we introduce a tailored stochastic model for the occurrence of crack initiations.

\section{Spatial Poisson process to model crack initiation} \label{Poisson}
For a given S-N model that depends on parameters $\theta$, we consider the following notations:
\begin{itemize}
\item The failure density function is denoted by $f_{SN}(n;s,\theta)$, 
\item The cumulative distribution function is denoted by $F_{SN}(n;s,\theta)$, 
\item The failure/hazard rate function is denoted by $h_{SN}(n;s,\theta)$, 
\end{itemize}

where $s$ is a generic symbol for the stress. To construct a spatial Poisson process for crack initiation in a domain $D$, we assume that 
\begin{enumerate}
\item[(a)] The (averaged) effective stress at $\bold{x} \in \partial D$ determines the crack formation at $\bold{x}$.
\item[(b)] Different cracks initiate independently.
\end{enumerate}

The spatial Poisson process is governed by the intensity function $ \lambda(\bold{x},n) \geq 0$, which relates the spatial location to the number of cycles, $n$. This intensity function depends on the effective stress, $\lambda(\bold{x},n) = \eta(n ; \sigma^{\Delta}_{\mathrm{eff}} (\bold{x}))$, where $\eta$ is a failure-rate function. The number of cracks occurring inside any surface region, $B \subset \partial D$, can be modeled as a Poisson counting process with the associated rate function $\lambda _B(n) = \int_{B}\lambda(\bold{x}, n)dS(\bold{x})$. Then, the number of crack initiations in the region $B$ after performing $n$ stress cycles is modeled as a Poisson random variable, $M_B(n)$, whose distribution is
\[ P(M_B(n) = m) = \frac{(\Lambda_B(n))^m}{m!}\exp(-\Lambda_B(n)) \, , m = 0, 1, \ldots \]
with $\Lambda_B(n) = \int_{0}^{n} \lambda_B(u) du \geq 0$.

We are interested in the number of stress cycles, $N_{\partial D}$, when the first crack initiates on $\partial D$. The random variable $N_{\partial D}$ is related to the counting process $M_{\partial D}(n)$ by the equivalence relation $N_{\partial D} > n \iff M_{\partial D}(n) = 0$, which corresponds to the survival event. Similarly, the failure event, $N_{\partial D} \leq n$, is equivalent to $M_{\partial D}(n) \geq 1$. Therefore, the survival probability after $n$ cycles is
\begin{eqnarray}
P(N_{\partial D} > n;\sigma^{\Delta}_{\mathrm{eff}}) &=& P(M_{\partial D}(n) = 0) = \exp(-\Lambda_{\partial D}(n))  \nonumber \\
&=& \exp\left(-\int_{0}^{n} \int_{\partial D} \eta(\hat{n}; \sigma^{\Delta}_{\mathrm{eff}}(\bold{x})) dS(\bold{x}) \, d\hat{n} \right), \label{surveq}
\end{eqnarray}
and the density function of $N_{\partial D}$ is
\begin{eqnarray*}
\rho^{\partial D}(n;\sigma^{\Delta}_{\mathrm{eff}}) &=& \frac{\partial}{\partial n} P(N_{\partial D} \leq n;\sigma^{\Delta}_{\mathrm{eff}}) \\
&=& \frac{\partial}{\partial n} \left( 1 - P(N_{\partial D} > n;\sigma^{\Delta}_{\mathrm{eff}}) \right) \\
&=& \frac{\partial}{\partial n} \left( 1 -  \exp\left(-\int_{0}^{n} \int_{\partial D} \eta(\hat{n}; \sigma^{\Delta}_{\mathrm{eff}}(\bold{x})) dS(\bold{x}) \, d\hat{n} \right) \right) \,\,\,\,\,\, (\text{from \eqref{surveq}}) \\
&=&  - \exp\left(-\int_{0}^{n} \int_{\partial D} \eta(\hat{n}; \sigma^{\Delta}_{\mathrm{eff}}(\bold{x})) dS(\bold{x}) \, d\hat{n} \right) \times \frac{\partial}{\partial n} \left( -\int_{0}^{n} \int_{\partial D} \eta(\hat{n}; \sigma^{\Delta}_{\mathrm{eff}}(\bold{x})) dS(\bold{x}) \, d\hat{n} \right) \\
&=& - P(N_{\partial D} > n;\sigma^{\Delta}_{\mathrm{eff}}) \times \frac{\partial}{\partial n} \left( -\int_{0}^{n} \int_{\partial D} \eta(\hat{n}; \sigma^{\Delta}_{\mathrm{eff}}(\bold{x})) dS(\bold{x}) \, d\hat{n} \right) \\
&=& P(N_{\partial D} > n;\sigma^{\Delta}_{\mathrm{eff}}) \times \int_{\partial D} \eta(n; \sigma^{\Delta}_{\mathrm{eff}}(\bold{x})) dS(\bold{x}).
\end{eqnarray*}

To parameterize the rate function $\eta(n; s)$, we relate it to a given S-N model as follows:
\begin{equation}
\label{etafun}
\eta(n; s) = -\frac{1}{\gamma} \frac{\partial}{\partial n} \log\left(1-F_{SN}(n; s,\theta) \right) = \frac{1}{\gamma}\frac{f_{SN}(n;s,\theta)}{1-F_{SN}(n;s,\theta)} = \frac{1}{\gamma} h_{SN}(n; s,\theta),
\end{equation}
where $\gamma$ is the size of the highly stressed volume. Thus, the spatial Poisson model is fully characterized by $\theta, \gamma$, and $\Delta$, where $\theta$ depends on the selected S-N model. 

From \eqref{etafun}, the survival probability can be simplified to
\begin{equation}
\label{surveq2}
P(N_{\partial D} > n;\sigma^{\Delta}_{\mathrm{eff}}) = \exp \left( \frac{1}{\gamma} \int_{\partial D} \log(1 - F_{SN}(n;\sigma^{\Delta}_{\mathrm{eff}}(\bold{x}), \theta)) dS(\bold{x}) \right).
\end{equation}

\begin{remark}
In the case of uniform stress, $\sigma^{\Delta}_{\mathrm{eff}}(\bold{x}) \equiv \sigma_{eq}$ and $\gamma = |\partial D|$. Thus, the survival probability \eqref{surveq2} becomes 
\[ P(N_{\partial D} > n; \sigma^{\Delta}_{\mathrm{eff}})  =  1 - F_{SN}(n;\sigma_{eq}, \theta) \,,\]
which means that the parameterization of $\eta(n; s)$ in \eqref{etafun} ensures the consistency of the model for uniform and non-uniform stresses. 
\end{remark}

The highly stressed volume (area), $\gamma$, depends on the specimen geometry. We re-parametrize the survival probability by defining the highly stressed volume. 

\begin{definition}\textbf{Highly stressed volume}\\
We let $\mathcal{A}_{\beta} = \{\bold{x} \in \partial D: \sigma^{1}_{\mathrm{eff}}(\bold{x}) > \beta\}$, where $\sigma^{1}_{\mathrm{eff}}(\bold{x})$ is the effective stress that corresponds to a unity traction and $\beta$ is an unknown parameter. The highly stressed volume is given by
\begin{equation}
\label{hsv}
\gamma(\beta) = \int_{\partial D}\mathbbm{1}_{\mathcal{A}_{\beta}}(\bold{x}) dS(\bold{x}).
\end{equation}
\end{definition}

Under the assumption of independent experiments, the log-likelihood function is

\[\ell(\theta, \beta, \Delta) = \sum_{i = 1}^{m} \left[ (1-\delta_{i}) \log( P(N_{\partial D} > n_{i}; \sigma^{\Delta}_{\mathrm{eff},i})) + \delta_i \log(\rho^{\partial D}(n_i; \sigma^{\Delta}_{\mathrm{eff},i})) \right] \]

\[ = \sum_{i=1}^{m} \left\{ (\delta_{i}-1) \int_{0}^{n_{i}} \int_{\partial D} \eta(\hat{n}; \sigma^{\Delta}_{\mathrm{eff},i}(\bold{x})) dS(\bold{x}) d\hat{n} +  \delta_i \log\left( \int_{\partial D} \eta(n_i; \sigma^{\Delta}_{\mathrm{eff},i}(\bold{x})) dS(\bold{x})\right) \right.\] \[ \left. - \delta_i \int_{0}^{n_{i}} \int_{\partial D} \eta(\hat{n}; \sigma^{\Delta}_{\mathrm{eff},i}(\bold{x})) dS(\bold{x}) d\hat{n} \right\}\]

\[ = \sum_{i=1}^{m} \left\{ (1-\delta_{i}) \frac{1}{\gamma(\beta)}\int_{\partial D} \log (1 - F_{SN}(n_{i}; \sigma^{\Delta}_{\mathrm{eff},i}(\bold{x}),\theta) ) dS(\bold{x}) +  \delta_i \log\left( \frac{1}{\gamma(\beta)} \int_{\partial D} h_{SN}(n_i, \sigma^{\Delta}_{\mathrm{eff},i}(\bold{x}), \theta)dS(\bold{x})\right) \right. \]  \[  \left. + \delta_i \frac{1}{\gamma(\beta)}\int_{\partial D} \log( 1 - F_{SN}(n_i; \sigma^{\Delta}_{\mathrm{eff},i}(\bold{x}),\theta)) dS(\bold{x}) \right\} \]

\begin{equation}
\label{loglike}
= \sum_{i=1}^{m} \left\{ \frac{1}{\gamma(\beta)}\int_{\partial D} \log ( 1- F_{SN}(n_i; \sigma^{\Delta}_{\mathrm{eff},i}(\bold{x}),\theta)) dS(\bold{x}) +  \delta_i \log\left( \frac{1}{\gamma(\beta)} \int_{\partial D} h_{SN}(n_i, \sigma^{\Delta}_{\mathrm{eff},i}(\bold{x}), \theta) dS(\bold{x})\right) \right\} .
\end{equation}

\subsection{Calibration of the spatial Poisson model}

Again, we consider Model Ia and calibrate the likelihood \eqref{loglike} under the assumption that $\Delta = 0$. We replace the averaged effective stress, $\sigma^{\Delta}_{\mathrm{eff}}(\bold{x})$ in the log-likelihood function \eqref{loglike} with the effective stress, $\sigma_{\mathrm{eff}}(\bold{x})$. The results of the maximization of the log-likelihood with respect to $\theta$ and $\beta$ are summarized in Table \ref{mle1hsv}. The overall fit in Table \ref{mle1hsv} is better than the fit in Table \ref{mle1anotch}. When moving from 2D to 3D modeling, we did not obtain any significant gain concerning the goodness of fit (therefore, at this stage, 2D modeling is recommended). We still notice variations in the estimated parameters with each data set as in Table \ref{mle1anotch}. However, the standard deviation $\tau$ is reduced considerably in all cases. Also, the difference between the calibrated fatigue-limit parameters from the 2D and 3D models of specimen 3 is small in this case. Therefore, we use only the 2D model for all three specimens from now on. We try to improve the fitting by incorporating $\Delta$.

\begin{table}[!htb]
\begin{center}
\caption{ML estimates of $\theta$ and $\beta$ from Model Ia using \eqref{loglike} where $\Delta = 0$.}
\begin{tabular}{|c|c|c|c|c|c|c|c|c|}
\hline
Data set & $A_1$ & $A_2$ & $A_3$ & $q$ & $\tau$ & $\beta$ & Maximum log-likelihood & AIC \\
\hline
1 & 5.88 & -1.32 & 35.88 &  0.5640 & 0.3011 & 1.16 & -938.90  &  1889.80 \\
\hline
2 & 6.00 &  -1.22  &  40.98  &  0.5965  &  0.2299 & 1.95  & -391.45 &  794.90 \\
\hline
3 (2D)  & 7.62 &  -2.18  &  45.35  &  0.6504  &  0.2692 &  2.54  & -301.82 &  615.64\\
\hline
3 (3D) & 7.62 &  -2.16  &  44.62  &  0.6504  &  0.2903 &  2.90  & -302.03 &  616.06 \\
\hline
1, 2 \& 3 (2D) & 6.28 &  -1.47 &  35.99  &  0.5676  &  0.3804 &  1.83  & -1650.05 &  3312.10 \\
\hline
1, 2 \& 3 (3D) & 6.30 &  -1.47  &  35.78  &  0.5646  &  0.3718  &  1.83  & -1649.97 & 3311.94 \\
\hline
\end{tabular}
\label{mle1hsv}
\end{center}
\end{table}

\begin{table}[h!]
\begin{center}
\caption{ML estimate of $\theta$, $\beta$ and $\Delta$ from Model Ia using \eqref{loglike}.}
\begin{tabular}{|c|c|c|c|c|c|c|c|c|c|}
\hline
Data set & $A_1$ & $A_2$ & $A_3$ & $q$ & $\tau$ & $\beta$ & $\Delta$ & \begin{tabular}{@{}c@{}}Maximum \\ log-likelihood \end{tabular} & AIC \\
\hline
1, 2 \& 3 (2D) & 6.29 &  -1.47  &  35.99  &  0.5664  &  0.3453  &  1.83  & 0.0125 & -1648.16 & 3310.32 \\
\hline
\end{tabular}
\label{mle1hsvdelta}
\end{center}
\end{table}

\begin{figure}[h!]
\centering
\includegraphics[width=9cm]{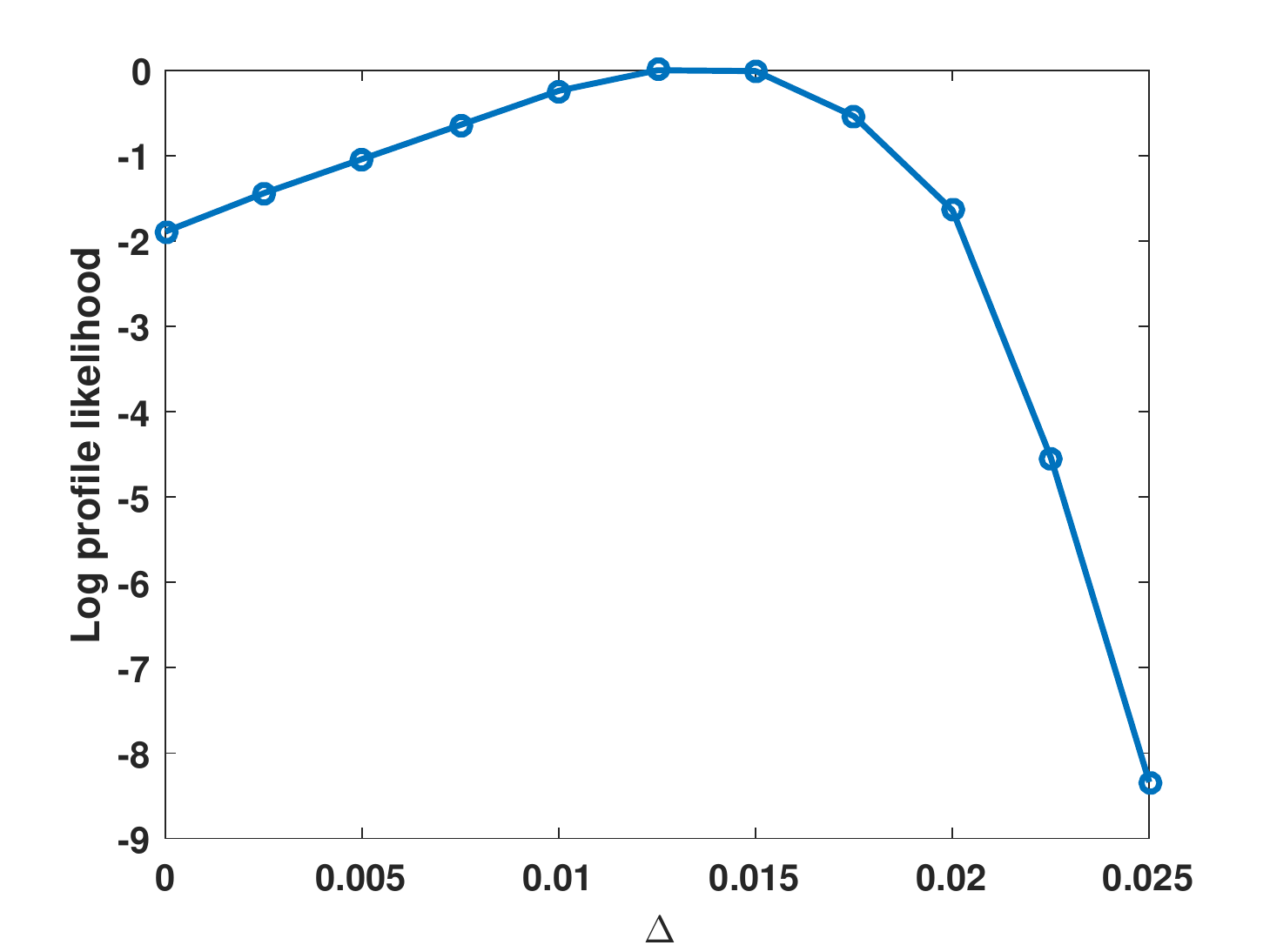}
\caption{The estimated log profile likelihood function for the parameter $\Delta$ derived from \eqref{loglike} and the combined data sets.}
\label{logprofLike2}
\end{figure}

The pooled ML estimates of $\theta$, $\beta$, and $\Delta$ is presented in Table \ref{mle1hsvdelta}. Figure \ref{logprofLike2} shows the log profile likelihood function for $\Delta$. We compare the case when $\Delta$ is neglected with the case when $\Delta$ is incorporated in terms of their AIC in Table \ref{AIC2}. Slight preference is given to Model Ia when $\Delta>0$ is calibrated with the other parameters. This slight preference may not justify the extra computations needed to estimate $\Delta$. However, it should pointed out that for specimens with diverse geometries, it is expected that $\Delta$ will play a role in the fitting process. Therefore, $\Delta$ must be calibrated given the available data to take into account the microstructural features. We further study this comparison using a Bayesian framework in the next section.

Comparing AIC in Tables \ref{AIC1} and \ref{AIC2}, we see that the spatial Poisson process provides much better fit than the model introduced in \ref{calb}, whether $\Delta = 0$ or $\Delta > 0$. Although both models are generalizations using simple assumptions of the same S-N model (Model Ia), the spatial Poisson model incorporates the complete stress field and accounts for the highly stressed volume. We emphasize that we adopted the same resolution for both models in sections \ref{calb} and \ref{Poisson}. A detailed mesh sensitivity analysis for the Poisson model is provided in Subsection \ref{ConvA}.

\begin{table}[h!]
\begin{center}
\caption{Comparison between two different specifications of Model Ia using \eqref{loglike}.}
\begin{tabular}{|c|c|c|c|}
\hline
 Model Ia given data sets 1, 2 \& 3 (2D)  & $\Delta = 0$ & $\Delta > 0$ \\
\hline
Maximum log-likelihood &  -1650.05   &  -1648.16 \\
\hline
Akaike information criterion (AIC) &  3312.10  &  3310.32 \\
\hline
\end{tabular}
\label{AIC2}
\end{center}
\end{table}

\subsection{Survival functions}
\label{survsec}
\begin{figure}[!htb]
\centering
\includegraphics[width=18cm]{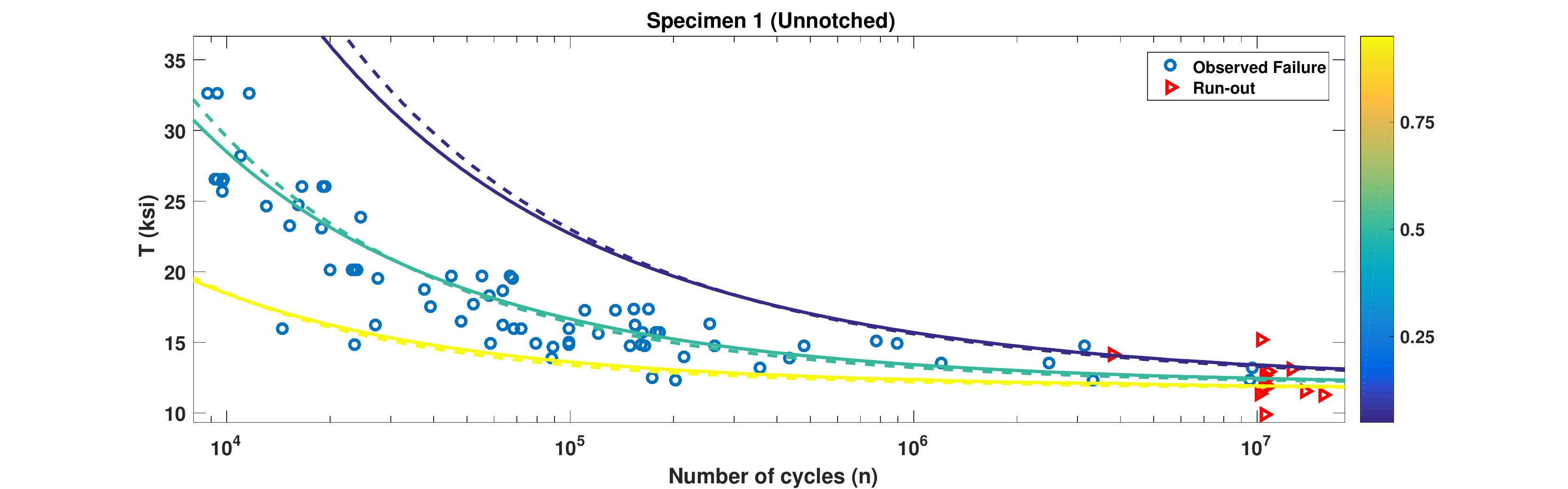}
\caption{Contour plots of the survival-probability functions for specimen 1 computed using \eqref{surveq2} with data set 1 ML estimates (dashed line) and pooled ML estimates (solid line); yellow is 0.95 probability, green is 0.5, and blue is 0.05.}
\label{PoissonC1}
\end{figure}

\begin{figure}[!htb]
\centering
\includegraphics[width=18cm]{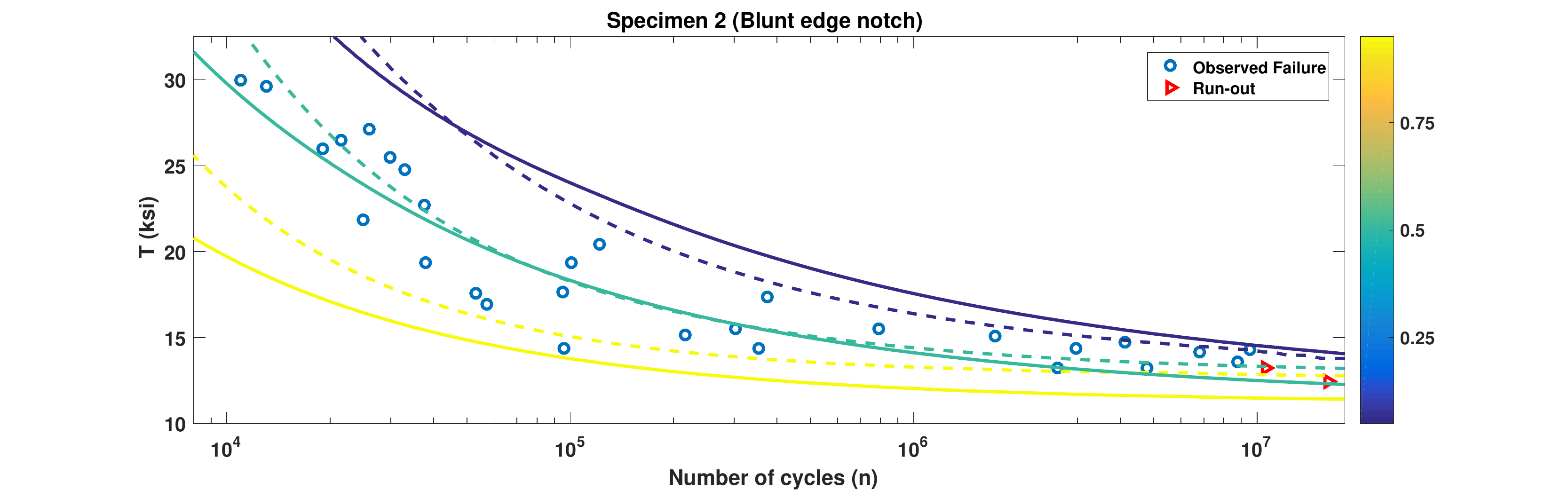}
\caption{Contour plots of the survival-probability functions for specimen 2 computed using \eqref{surveq2} with data set 2 ML estimates (dashed line) and pooled ML estimates (solid line); yellow is 0.95 probability, green is 0.5, and blue is 0.05.}
\label{PoissonC2}
\end{figure}

\begin{figure}[!htb]
\centering
\includegraphics[width=18cm]{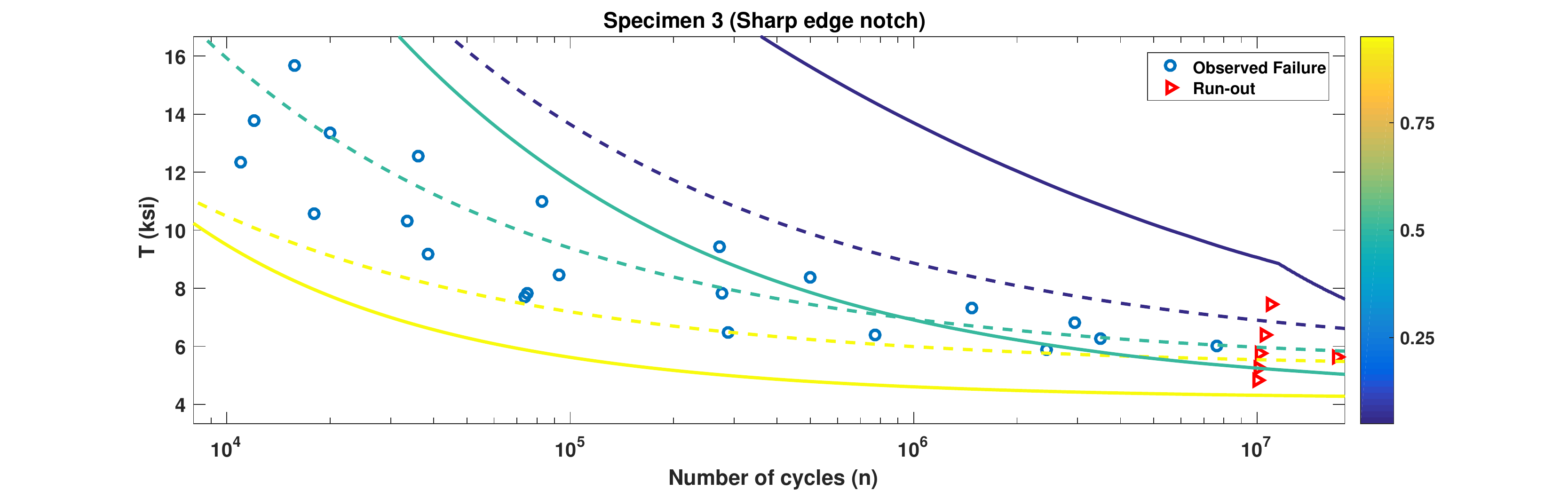}
\caption{Contour plots of the survival-probability functions for specimen 3 computed using \eqref{surveq2} with data set 3 ML estimates (dashed line) and pooled ML estimates (solid line); yellow is 0.95 probability, green is 0.5, and blue is 0.05.}
\label{PoissonC3}
\end{figure}

We compare the survival-probability functions for each specimen using model parameters estimated from each of their individual data sets as well as using the combined data set. The pooled ML estimates can be used to predict the survival probability of new specimens but with larger variability. The ML estimates obtained from the three data sets and the combined data set are provided in Table \ref{mle1hsv}. The survival-probability could be expressed as a function of the traction $T = \frac{W_{min}}{W_{max}} S_{max}\,(1-R)^{q}$ and the number of cycles $n$. We plot the contour lines of the survival-probability surface for each specimen together with its fatigue experiment data. Figures \ref{PoissonC1}, \ref{PoissonC2}, and \ref{PoissonC3} provide analog representations of the S-N curves computed using the Poisson model with the individual data sets estimates and the pooled estimates. In this case, the survival probability depends on the complete stress field of the specimen, and therefore, there are no unique S-N curves for the three specimens jointly.

\begin{figure}[!htb]
\centering
\includegraphics[width=18cm]{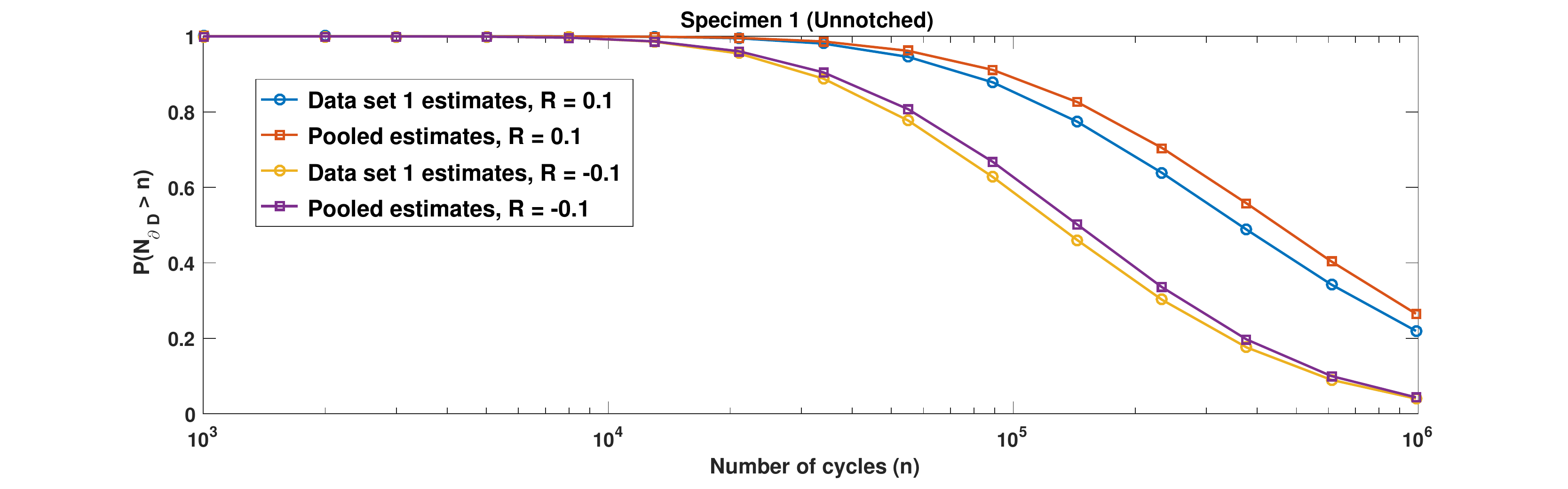}
\caption{Survival-probability functions for specimen 1 obtained using different ML estimates; $S_{max} = 45$ ksi.}
\label{surv1}
\end{figure}

\begin{figure}[!htb]
\centering
\includegraphics[width=18cm]{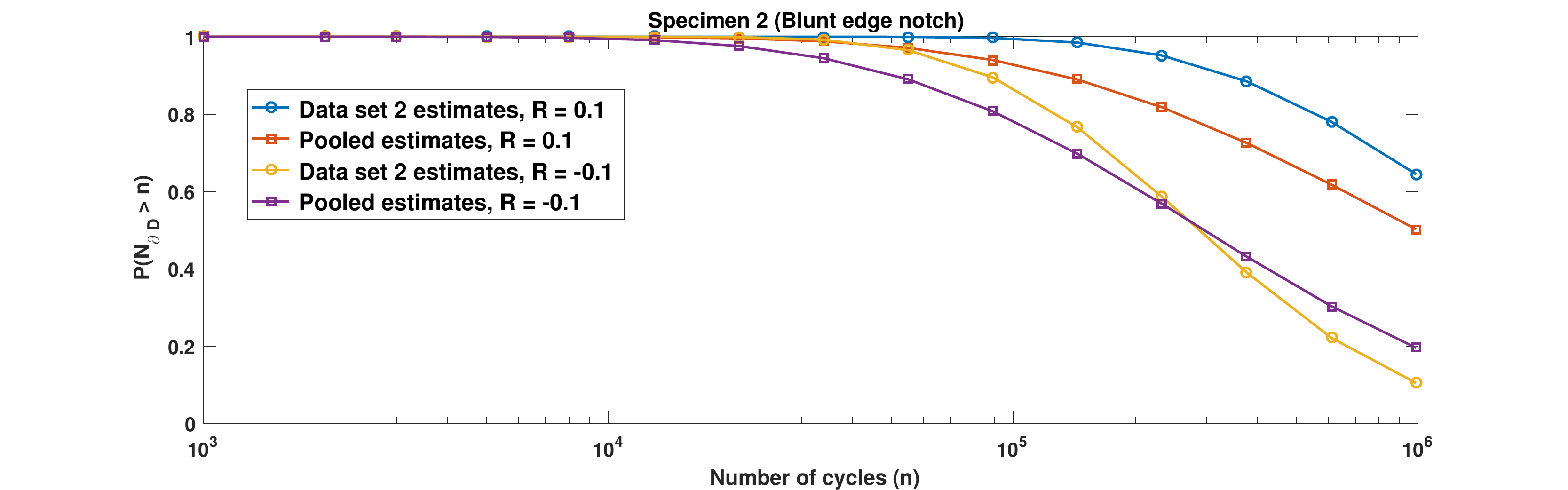}
\caption{Survival-probability functions for specimen 2 obtained using different ML estimates; $S_{max} = 30$ ksi.}
\label{surv2}
\end{figure}

\begin{figure}[!htb]
\centering
\includegraphics[width=18cm]{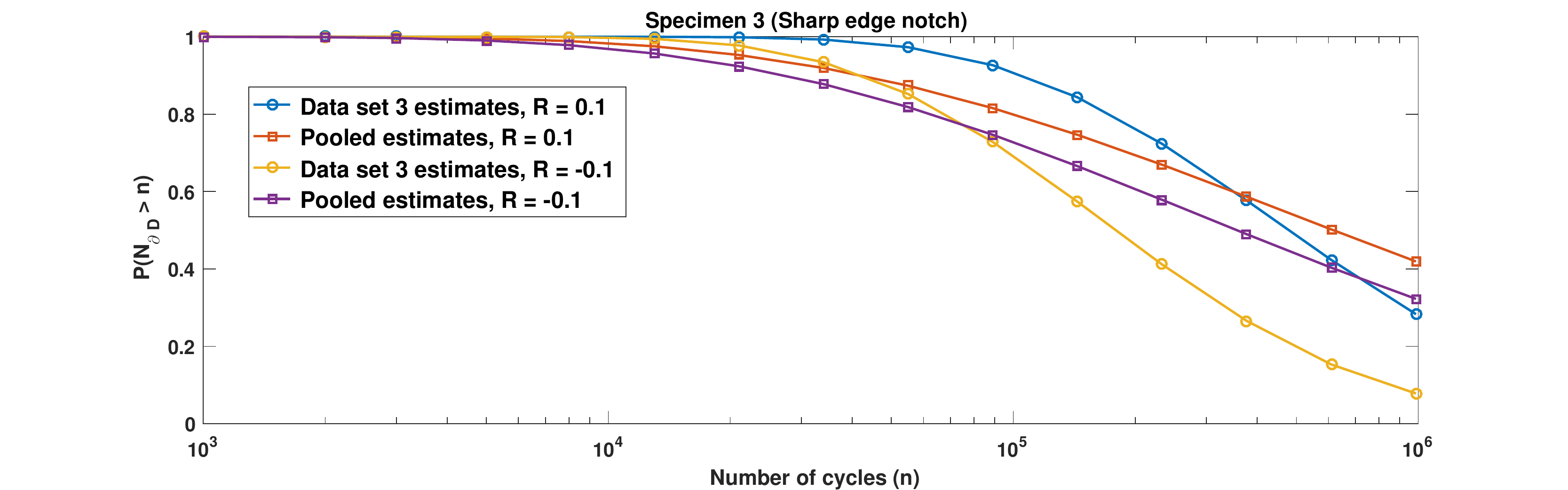}
\caption{Survival-probability functions for specimen 3 obtained using different ML estimates; $S_{max} = 12$ ksi.}
\label{surv3}
\end{figure}

In Figure \ref{PoissonC1}, we observe that the survival-probability functions computed from data set 1 estimates and the pooled estimates are similar. In Figure \ref{PoissonC2}, we notice a small increase in variability of the survival-probability function obtained by the pooled estimates compared. This variability increases considerably for specimen 3 in Figure \ref{PoissonC3}. However, the variability introduced by the pooled estimates provide more conservative survival functions that could be used for predictions and making design rules \cite{szabo2017formulation}. The variability increased with the pooled estimates because we assume spatial independence. In a single data set case, the spatial correlation influences all the estimated parameters, and therefore, the variance is reduced. We also compare the survival-probability functions for specific values of $S_{max}$ with $R = 0.1$ and $R = -0.1$ in Figures \ref{surv1}, \ref{surv2} and \ref{surv3}.

Figure \ref{surv1} shows the survival-probability functions for specimen 1, assuming $S_{max} = 45$ ksi. The survival-probability functions produced by the pooled estimates are very similar to the ones produced from estimates obtained using only the dataset 1. Including fatigue data from the notched specimens has only a small effect on the survival probability of the unnotched specimen. Figures \ref{surv2} and \ref{surv3} show the survival-probability functions for specimens 2 and 3 using $S_{max} = 30$ ksi and $S_{max} = 12$ ksi, respectively. The estimated survival of specimen 2 computed using the pooled estimates is more pessimistic than the estimated survival of specimen 2 using the data set 2 estimates. For specimen 3, the survival functions using the pooled estimates are initially smaller (more pessimistic) than the survival functions using the data set 3 estimates. However, the estimated survival functions from the data set 3 decay much faster with the number of cycles than those of the pooled estimates. Thus, the effect of pooling data sets is substantial for specimens 2 and 3. One possible reason is that data sets 2 and 3 are much smaller than data set 1. 

\subsection{Convergence analysis}\label{ConvA}

In this subsection, we study the effect of finite element triangulation on the estimated parameters and analyze the convergence of the ML estimates with respect to the mesh. To simplify the analysis, we consider the case of $\Delta = 0$. In Tables \ref{conv1t}, \ref{conv2t} and \ref{conv3t}, we present the ML estimates of $\theta$ and $\beta$ for specimens 1, 2 and 3, respectively, under different meshes. 

\begin{table}[!htb]
\begin{center}
\caption{ML estimates for Model Ia given data set 1 with different meshes for specimen 1.}
\begin{tabular}{|c|c|c|c|c|c|c|c|}
\hline
\# triangles & $A_1$ & $A_2$ & $A_3$ & $q$ & $\tau$ & $\beta$ & Maximum log-likelihood \\
\hline
60 & 5.77 & -1.23 & 35.54 &  0.5611 & 0.3112 & 1.16 & -939.07 \\
\hline
593 & 5.88 & -1.32 & 35.84 &  0.5631 & 0.2975 & 1.14 & -939.08 \\
\hline
1275 & 5.88 & -1.31 & 35.87 &  0.5636 & 0.3004 & 1.16 & -938.90 \\
\hline
2852 & 5.88 & -1.32 & 35.87 &  0.5646 & 0.3020 & 1.16 & -938.92 \\
\hline
9948 & 5.88 & -1.32 & 35.88 &  0.5640 & 0.3011 & 1.16 & -938.90 \\
\hline
\end{tabular}
\label{conv1t}
\end{center}
\end{table}

\begin{table}[!htb]
\begin{center}
\caption{ML estimates for Model Ia given data set 2 with different meshes for specimen 2.}
\begin{tabular}{|c|c|c|c|c|c|c|c|}
\hline
\# triangles & $A_1$ & $A_2$ & $A_3$ & $q$ & $\tau$ & $\beta$ & Maximum log-likelihood \\
\hline
74 & 5.39 & -0.0514 &  24.98  &  0.5496  &  0.5640  &  1.7  &  -395.50 \\
\hline
611 & 6.00 & -1.25 &  39.48  &  0.5890  &  0.2395  &  1.83  &  -391.62 \\
\hline
1294 & 6.00 & -1.21 &  40.54  &  0.5994  &  0.2358  &  1.98  &  -391.62 \\
\hline
2580 & 6.00 & -1.23 &  40.79  &  0.5926  &  0.2241  &  1.91  &  -391.24 \\
\hline
6508 & 6.00 &  -1.22  &  40.98  &  0.5965  &  0.2299 & 1.95  & -391.45 \\
\hline
\end{tabular}
\label{conv2t}
\end{center}
\end{table}

\begin{table}[!htb]
\begin{center}
\caption{ML estimates for Model Ia given data set 3 with different meshes for specimen 3 (2D).}
\begin{tabular}{|c|c|c|c|c|c|c|c|c|}
\hline
\# triangles & $A_1$ & $A_2$ & $A_3$ & $q$ & $\tau$ & $\beta$ & Maximum log-likelihood \\
\hline
352 & 7.14 & -1.83 & 37.15 &  0.6445 & 0.4134 & 3.66 & -301.50 \\
\hline
3840 & 7.40 & -2.04 & 44.60 &  0.6538 & 0.2797 & 2.62 & -302.03 \\
\hline
6664 & 7.5 & -2.10 & 45.11 &  0.6504 & 0.2680 & 2.53 & -301.83 \\
\hline
15649 &  7.62 &  -2.18  &  45.41  &  0.6524  &  0.2729 &  2.59  & -301.87 \\
\hline
26544 & 7.62 &  -2.18  &  45.35  &  0.6504  &  0.2692 &  2.54  & -301.82 \\
\hline
\end{tabular}
\label{conv3t}
\end{center}
\end{table}

\begin{figure}[!htb]
\minipage{0.32\textwidth}
\includegraphics[width=\linewidth]{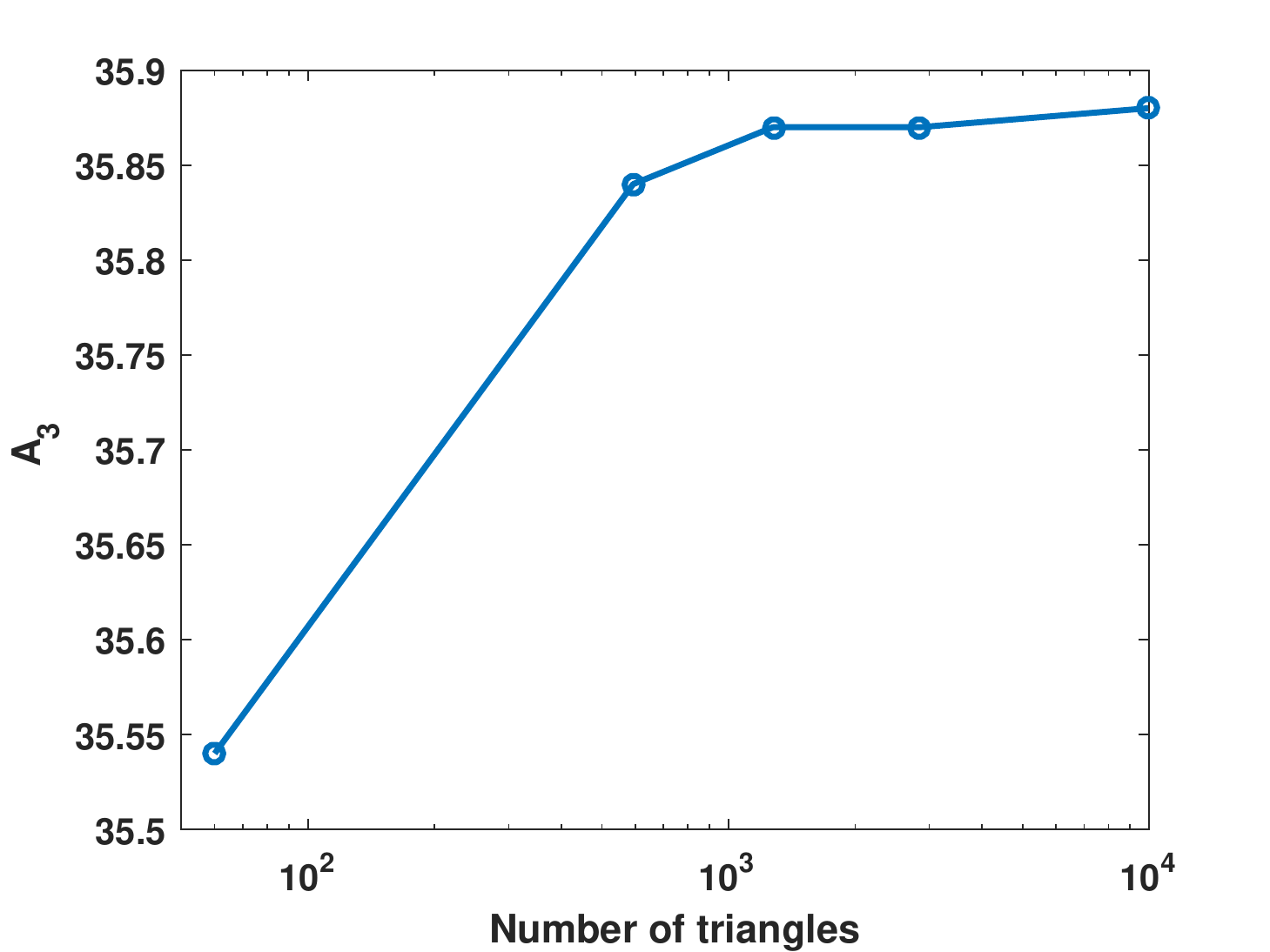}
\endminipage\hfill
\minipage{0.32\textwidth}
\includegraphics[width=\linewidth]{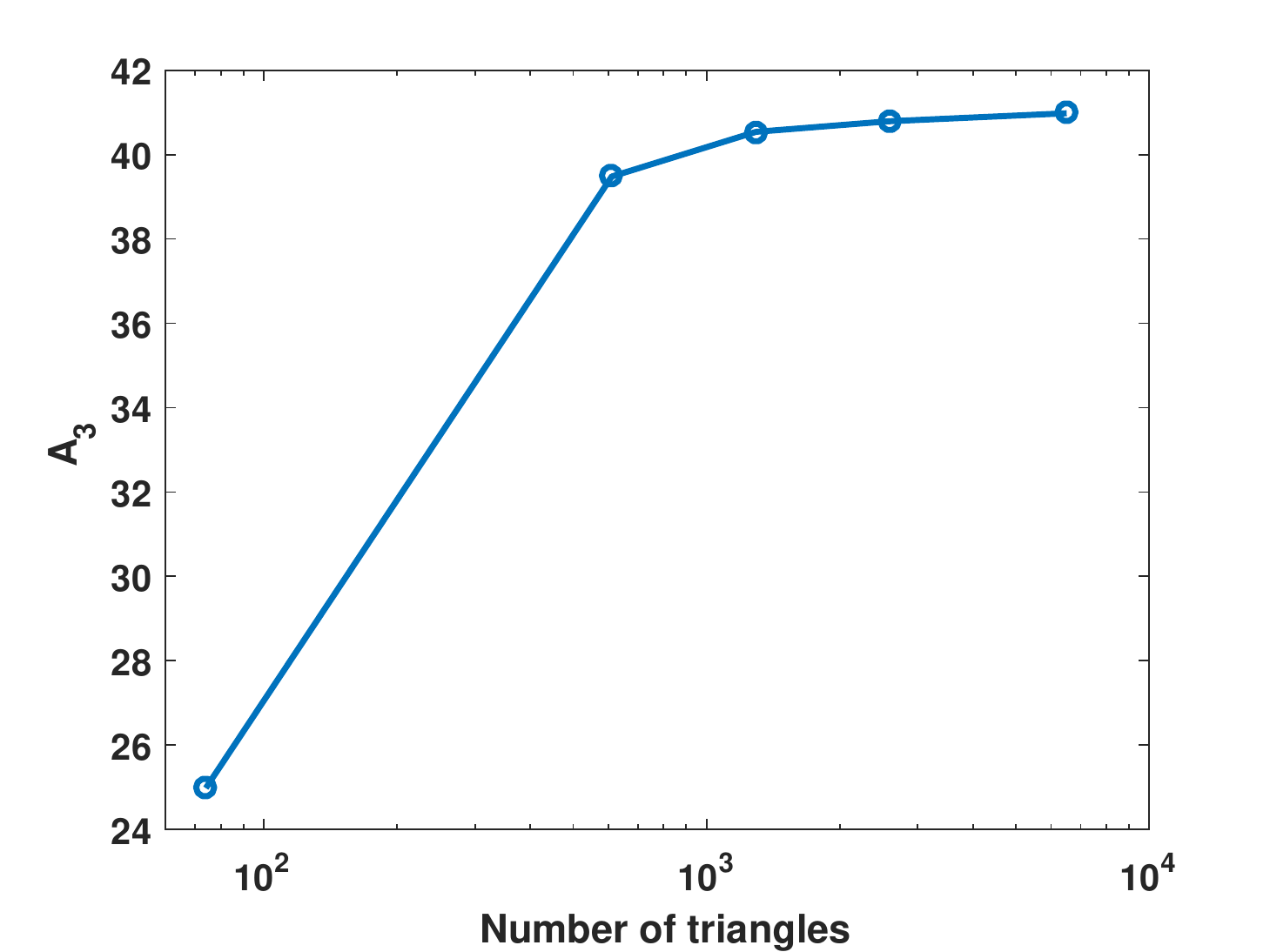}
\endminipage\hfill
\minipage{0.32\textwidth}
\includegraphics[width=\linewidth]{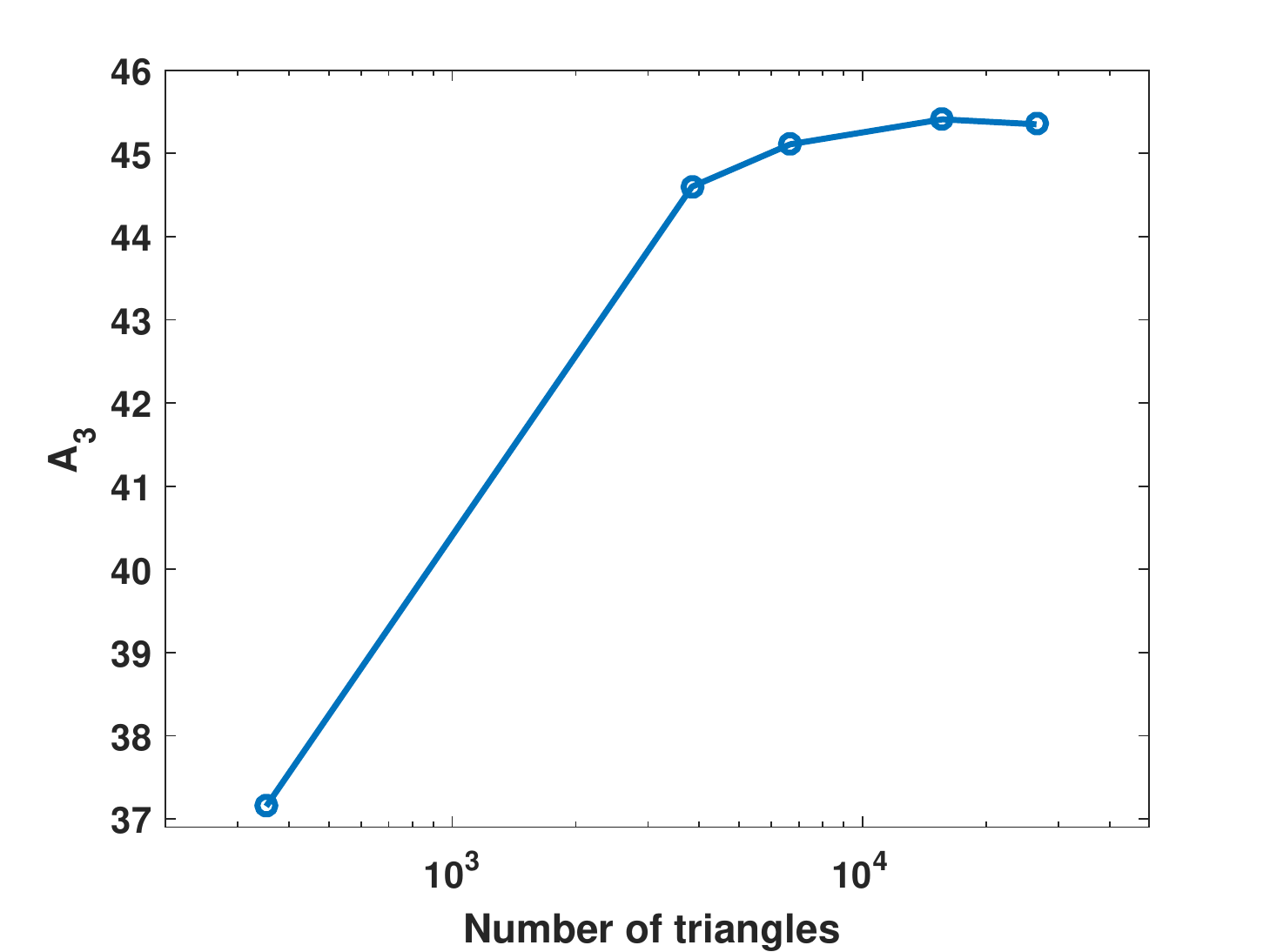}
\endminipage
\caption{Convergence of the ML estimates of the fatigue-limit parameter $A_3$ with the number of triangles used for specimen 1 (left panel), specimen 2 (middle panel), and specimen 3 (right panel).}
\label{conv123}
\end{figure}

Figure \ref{conv123} shows the convergence of the estimated fatigue-limit parameter, $A_3$ with respect to the number of triangles used in the finite element triangulation. The other estimated parameters exhibit similar behavior. This analysis also ensures the robustness of the estimated survival functions in \ref{survsec}.

\section{Bayesian analysis}
\label{Bayesian}
So far, we have analyzed the spatial Poisson model following a classical statistical approach. Now, we consider a Bayesian framework and infer the Model Ia parameters under the two scenarios of $\Delta = 0$ and $\Delta = 0.0125$. We assume the following uniform priors:
\[ A_1 \sim U(2, 13) , \, A_2 \sim U(-7, 0), \, A_3 \sim U(20, 40) , \, q \sim U(09.1, 1) , \, \tau \sim U(0.01, 1.5) , \, \beta \sim U(0.01, 5) \]

We run Markov chain Monte Carlo (MCMC) algorithm to generate samples from the joint posterior distribution using the combined data sets. We plot the marginal posterior distributions and compare the results of the two scenarios using the Bayes factor (log marginal likelihood) and the deviance information criterion (DIC). We refer the reader to \cite{fatigue1} for the details of the MCMC algorithm and the Bayesian comparison tools.

\begin{figure}[!htb]
\centering
\includegraphics[width=18cm]{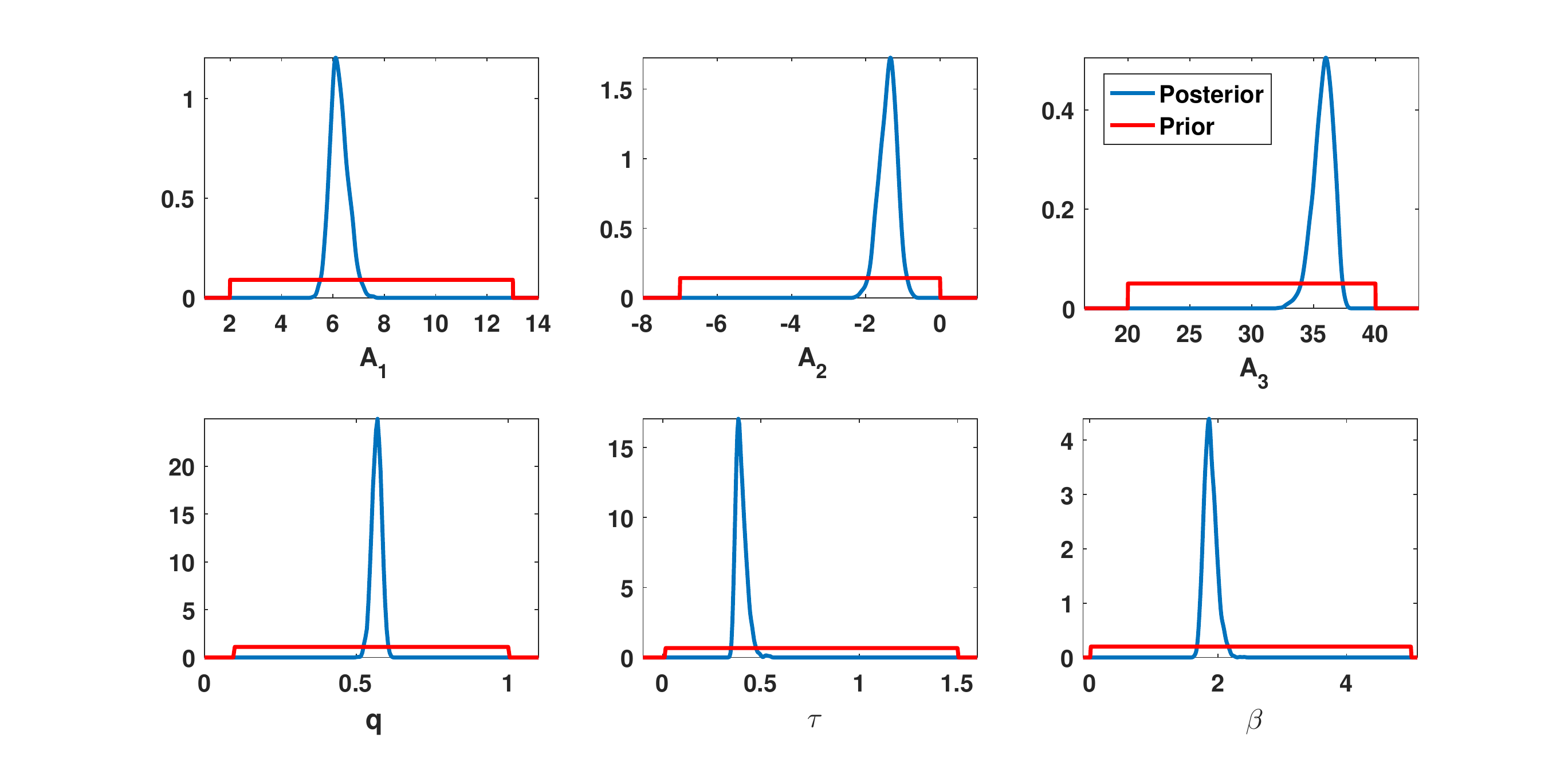}
\caption{The estimated marginal posteriors for the parameters $A_1, A_2, A_3, q, \tau, \beta$, where $\Delta = 0$.}
\label{MCMC1}
\end{figure} 

\begin{table}[!htb]
\begin{center}
\caption{The estimated correlation coefficients of each pair of the parameters $A_1, A_2, A_3, q, \tau, \beta$, where $\Delta = 0$.}
\begin{tabular}{|c|c|c|c|c|c|c|c|}
\hline
        & $A_1$ & $A_2$ & $A_3$ & $q$ & $\tau$ \\
\hline
$A_2$   & -0.98 & ---  & --- & ---   & --- \\
\hline
$A_3$   & -0.85 & 0.82 & --- & ---  & --- \\
\hline
$q$      & -0.43 & 0.42 & 0.41 & ---  & --- \\
\hline
$\tau$  & 0.27 & -0.23 & -0.25 &  -0.17  &  --- \\
\hline
$\beta$ & 0.43 & -0.30 & -0.19  &  -0.20  &  0.51 \\
\hline
\end{tabular}
\label{MCMCbiv1}
\end{center}
\end{table}

Figure \ref{MCMC1} shows the estimated marginal posterior distributions of the Model Ia parameters using the combined data sets and assuming $\Delta = 0$. We see that the data sets used are informative for all the parameters. The correlation coefficients are presented in Table \ref{MCMCbiv1}. The parameters $A_1$ and $A_2$ are highly correlated, and both are also correlated with the fatigue-limit parameter $A_3$. In general, the results obtained here are similar to the results obtained in \cite{fatigue1} using only the simple uniaxial Model Ia and data set 1.

\begin{figure}[!htb]
\centering
\includegraphics[width=18cm]{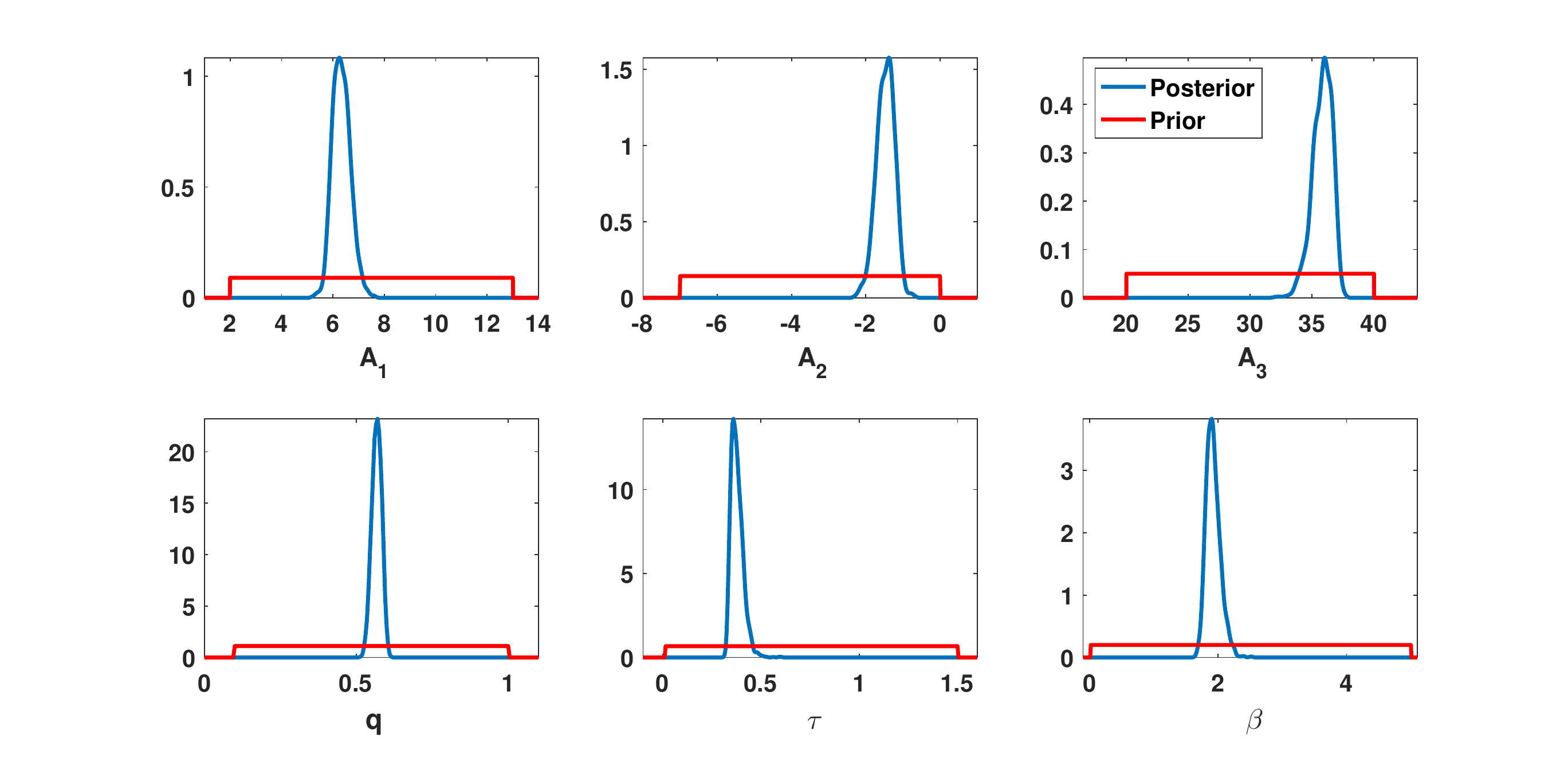}
\caption{The estimated marginal posteriors for the parameters $A_1, A_2, A_3, q, \tau, \beta$ where $\Delta = 0.0125$.}
\label{MCMC2}
\end{figure} 

\begin{table}[!htb]
\begin{center}
\caption{The estimated correlation coefficients of each pair of the parameters $A_1, A_2, A_3, q, \tau, \beta$, where $\Delta = 0.0125$.}
\begin{tabular}{|c|c|c|c|c|c|c|c|}
\hline
        & $A_1$ & $A_2$ & $A_3$ & $q$ & $\tau$ \\
\hline
$A_2$   & -0.97 & ---  & --- & ---   & --- \\
\hline
$A_3$   & -0.86 & 0.84 & --- & ---  & --- \\
\hline
$q$      & -0.44 & 0.44 & 0.41 & ---  & --- \\
\hline
$\tau$  & 0.16 & -0.08 & -0.16 &  -0.06  &  --- \\
\hline
$\beta$ & 0.37 & -0.22 & -0.14  &  -0.10  &  0.57 \\
\hline
\end{tabular}
\label{MCMCbiv2}
\end{center}
\end{table}

\begin{table}[!htb]
\begin{center}
\caption{Bayesian comparison between two different specifications of Model Ia using \eqref{loglike}.}
\begin{tabular}{|c|c|c|c|}
\hline
 Model Ia given data sets 1, 2 \& 3 (2D)  & $\Delta = 0$ & $\Delta = 0.0125$ \\
\hline
Log marginal likelihood &  -1660.24  &  -1658.48 \\
\hline
Deviance information criterion (DIC) &  3311.33  &  3308.27 \\
\hline
\end{tabular}
\label{DIC}
\end{center}
\end{table}

Figure \ref{MCMC2} and Table \ref{MCMCbiv2} show the estimated marginal posterior distributions and the correlation coefficients, respectively, when $\Delta = 0.0125$. The results are almost identical to the previous case, except for the parameter $\tau$, which now has a slightly smaller mode. We approximate the log marginal likelihood by the Laplace-Metropolis estimator \cite{lewraf} and compute DIC using the MCMC posterior samples. Table \ref{DIC} shows that Model Ia performs better with $\Delta = 0.0125$ than with $\Delta = 0$. However, the small difference between the two cases suggests that conclusions cannot be generalized to other S-N models. 

\subsection{Survival functions}
\label{SurvBayesian}

In this subsection, we compute some of the survival functions presented in \ref{survsec}, this time using the MCMC posterior samples of the model parameters obtained in the previous section with $\Delta = 0$, instead of the pooled ML estimates. Also, we consider a ``reference'' survival function that is obtained from the ML estimate of each data set.

\begin{figure}[!htb]
\centering
\includegraphics[width=18cm]{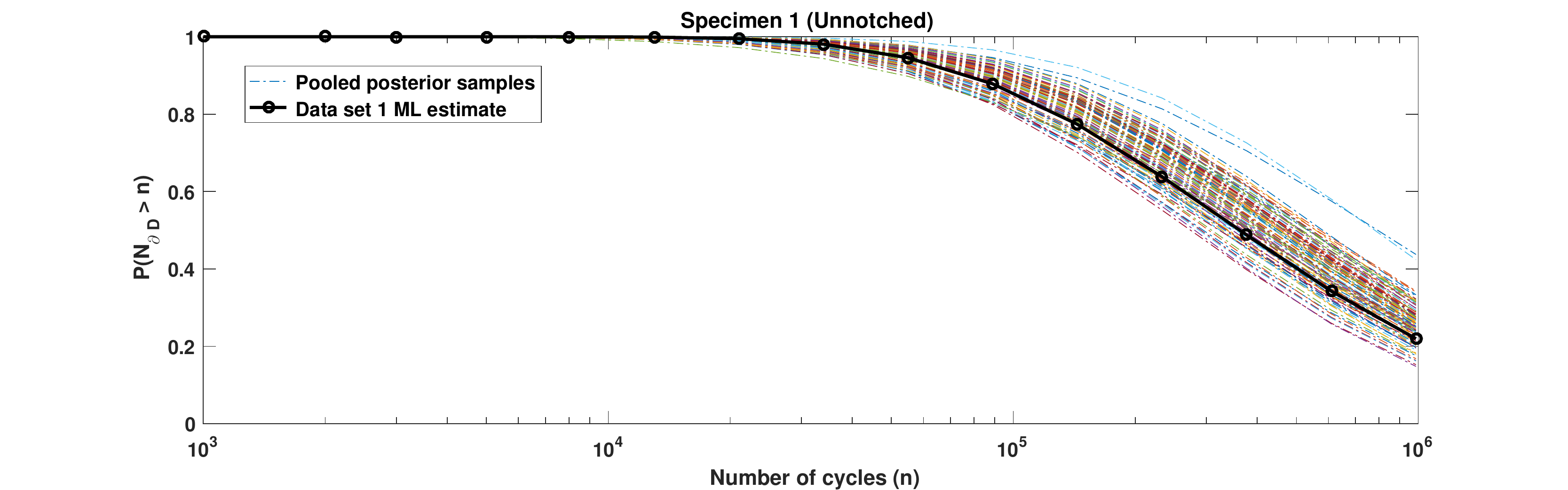}
\caption{Survival-probability functions for specimen 1 when $S_{max} = 45$ ksi and $R = 0.1$, given different posterior samples of the parameters $A_1, A_2, A_3, q, \tau, \beta$.}
\label{surv1mcmc}
\end{figure} 

\begin{figure}[!htb]
\centering
\includegraphics[width=18cm]{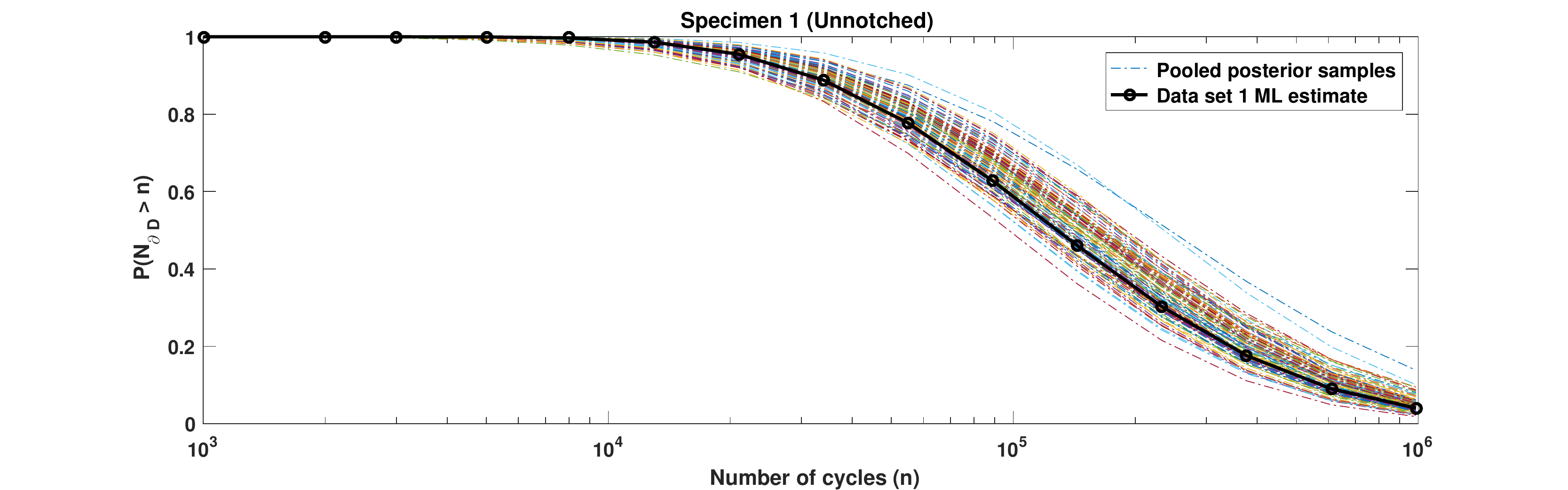}
\caption{Survival-probability functions for specimen 1 when $S_{max} = 45$ ksi and $R =- 0.1$, given different posterior samples of the parameters $A_1, A_2, A_3, q, \tau, \beta$.}
\label{surv1mcmc2}
\end{figure} 

\begin{figure}[!htb]
\centering
\includegraphics[width=18cm]{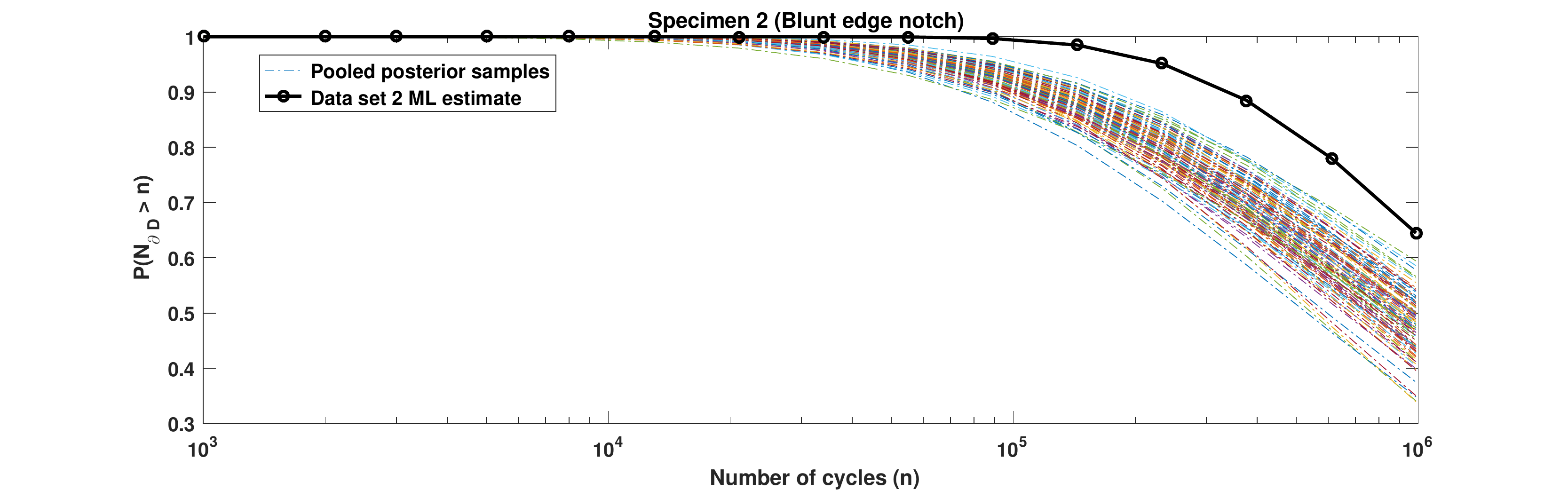}
\caption{Survival-probability functions for specimen 2 when $S_{max} = 30$ ksi and $R = 0.1$, given different posterior samples of the parameters $A_1, A_2, A_3, q, \tau, \beta$.}
\label{surv2mcmc}
\end{figure}

\begin{figure}[!htb]
\centering
\includegraphics[width=18cm]{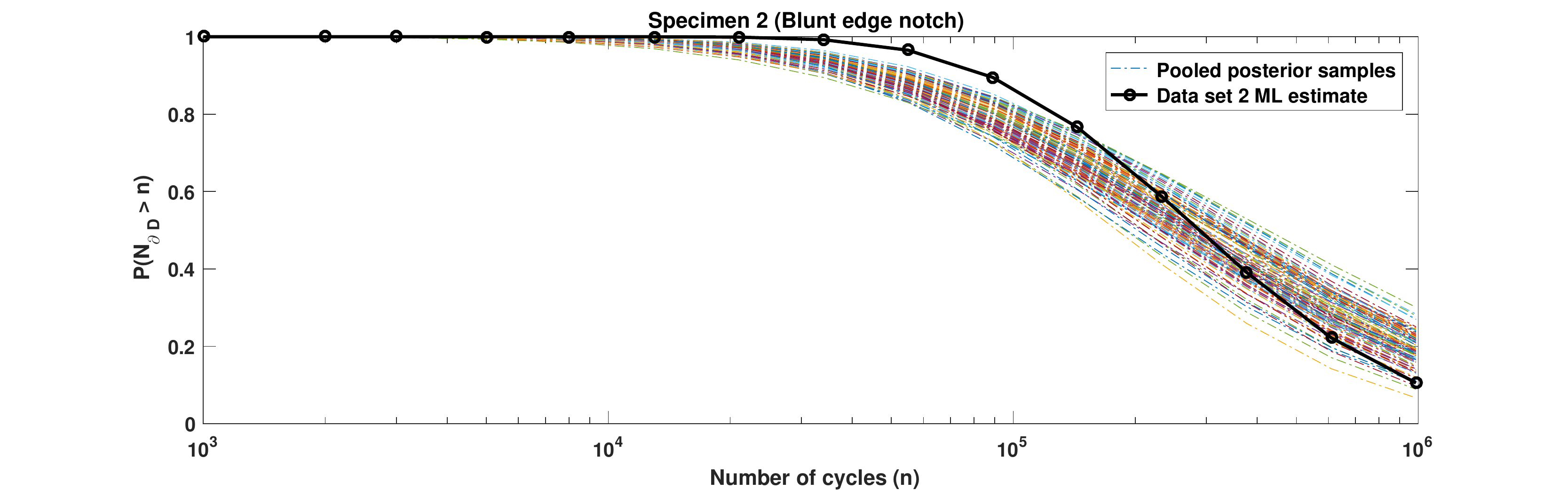}
\caption{Survival-probability functions for specimen 2 when $S_{max} = 30$ ksi and $R = -0.1$, given different posterior samples of the parameters $A_1, A_2, A_3, q, \tau, \beta$.}
\label{surv2mcmc2}
\end{figure}

Figures \ref{surv1mcmc} and \ref{surv1mcmc2} show the survival functions for specimen 1 when $R = 0.1$ and $R = -0.1$, respectively, and $S_{max} = 45$. The reference survival function falls within the cloud of survival functions obtained from the pooled posterior samples. Figures \ref{surv2mcmc} and \ref{surv2mcmc2} show the survival functions for specimen 2 when $S_{max} = 30$. The pooled posterior samples produce more conservative survival functions than the reference function. However, the difference between the survival functions cloud and the reference function is not significant.

\begin{figure}[!htb]
\centering
\includegraphics[width=18cm]{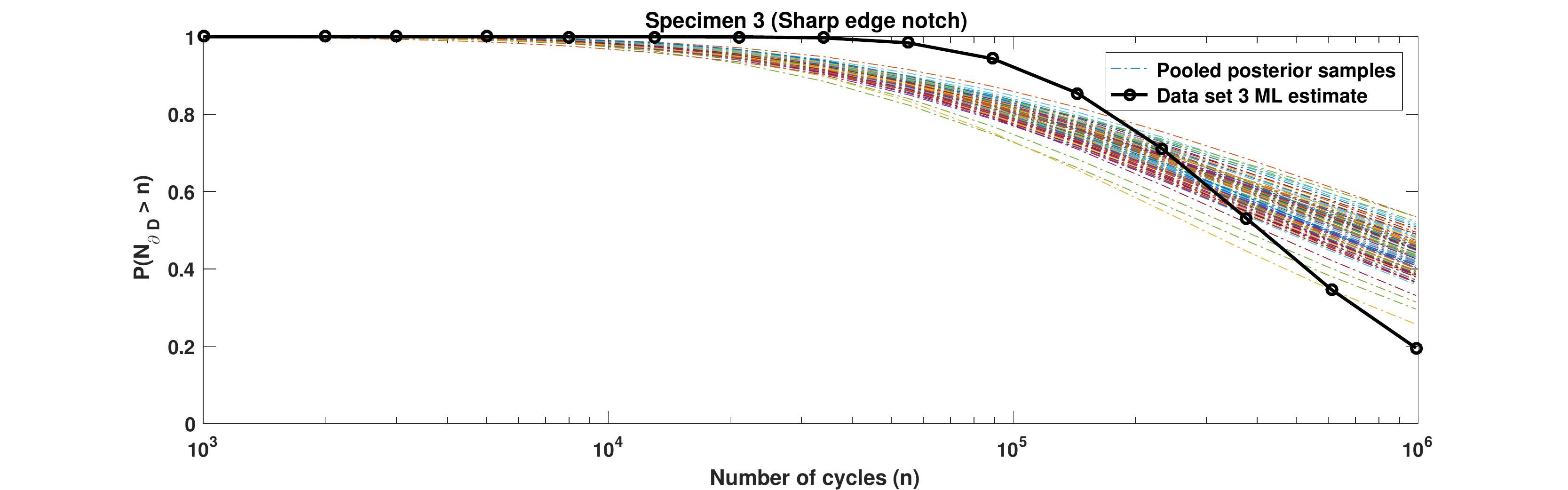}
\caption{Survival-probability functions for specimen 3 when $S_{max} = 12$ ksi and $R = 0.1$, given different posterior samples of the parameters $A_1, A_2, A_3, q, \tau, \beta$.}
\label{surv3mcmc}
\end{figure} 

\begin{figure}[!htb]
\centering
\includegraphics[width=18cm]{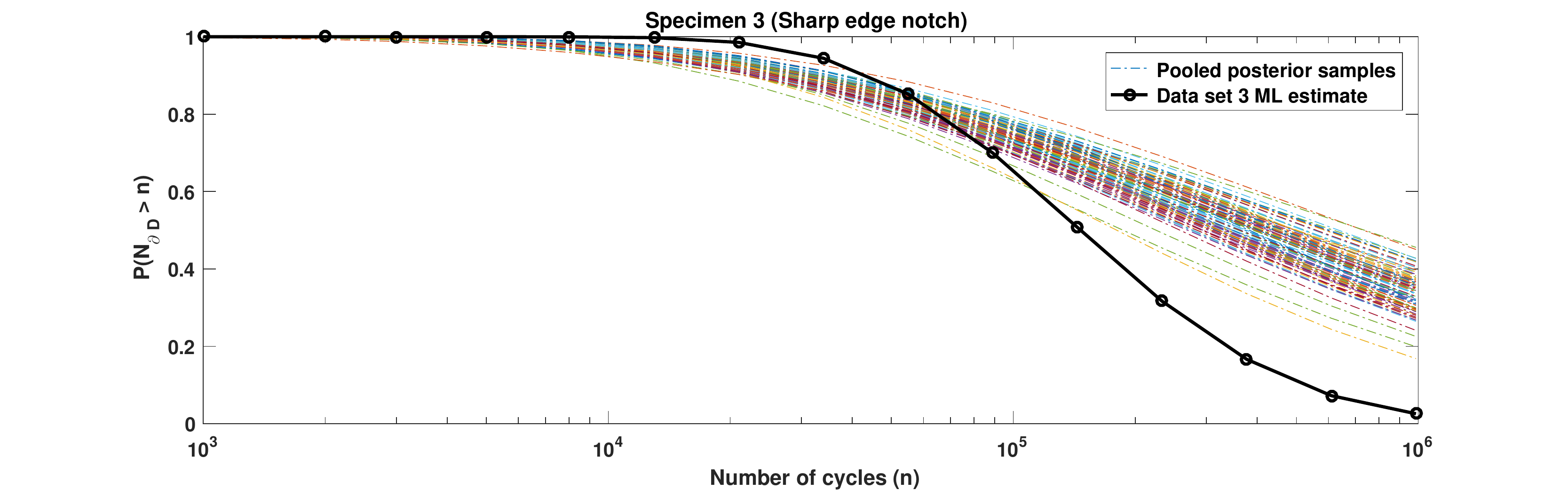}
\caption{Survival-probability functions for specimen 3 when $S_{max} = 12$ ksi and $R = -0.1$, given different posterior samples of the parameters $A_1, A_2, A_3, q, \tau, \beta$.}
\label{surv3mcmc2}
\end{figure} 

For specimen 3, we show the survival functions when $S_{max} = 12$ in Figures \ref{surv3mcmc} and \ref{surv3mcmc2}. For $R=0.1$, the pooled posterior samples provide more conservative survival probability than the reference function, especially when $n<2.5 \times 10^5$. For $R=-0.1$, the survival functions from the pooled sample are more optimistic than the reference function when $n>10^5$. In general, we notice that the survival functions are concentrated at high probability levels and that the dispersion increases at low probability levels. 

\section{Summary and conclusions}
The stress tensor is computed from the solution of the linear elasticity equations. A spatial stress function, $\sigma^{\Delta}_{\mathrm{eff}}(\bold{x})$, is defined by the maximum principal stress (where $\Delta = 0$) or by the local average of the maximum principal stress. The stress function can be combined with any statistical S-N curve without prior knowledge of the spatial correlation, under two ``extreme'' assumptions. The first assumption is the full spatial dependence on the global maximum value of the effective stress. Implicitly, this assumption means that the first crack will initiate at the position with the highest stress. The second assumption is complete spatial independence, which is modeled by the spatial Poisson process. For both assumptions, we calibrate the model parameters using the ML approach, given the data from the fatigue experiments. The simplifying assumptions are needed because we lack information about the spatial correlation. It is expected then to increase the variance when calibrating different data sets.

For simplicity, we first fit the models to the fatigue experiments on each specimen separately, assuming $\Delta = 0$. Then, we fit the models to the combined data set when $\Delta = 0$ and $\Delta > 0$. When calibrating $\Delta$, specimens with different geometries must be used for the parameter $\Delta$ to be identifiable. With the available data, the effect of $\Delta$ was limited. However, $\Delta$ is a fundamental parameter that must be calibrated. For specimen 3, we compare the effects of using biaxial stress (2D) and triaxial stress (3D) on fitting data set 3. When the likelihood function is given by \eqref{loglike} and the highly stressed volume is given by \eqref{hsv}, the difference is negligible. For more complicated geometries such as v-notched specimens, the 3D effect might be more prominent.

By classical model comparison tools, the spatial Poisson model is superior to the full spatial dependence assumption. Also, we find that the spatial Poisson model with $\Delta>0$ provides the best fit for the combined data sets. In the Bayesian framework, the estimated posteriors of the parameters are similar to the results obtained in \cite{fatigue1} with data set 1 except for the variance parameter $\tau$ that is reduced because we improved the model and added the new threshold parameter $\beta$ in the highly stressed volume. The deviance information criterion suggests that non-zero $\Delta$ improves slightly the fit of the combined data sets for Model Ia. To draw general conclusions, several S-N models must be studied and compared.

The spatial Poisson process provides a systematic approach to generalize uniaxial S-N models and calibrate fatigue experiments on different specimens without special treatments for notches. Given a sufficient number of fatigue experiments for specimens with diverse geometries, our proposed approach could be used to predict the life of any mechanical component made from the same material and having the same surface finish.
  
\section*{Acknowledgements}
Z. Sawlan, M. Scavino and R. Tempone are members of the KAUST SRI Center for Uncertainty Quantification in Computational Science and Engineering. R. Tempone received support from the KAUST CRG3 Award Ref: 2281 and the KAUST CRG4 Award Ref: 2584.

\appendix

\section{Matlab PDE solver}

The MATLAB function ADAPTMESH is used to solve the elliptic system PDE problem in 2D:

\begin{equation*}
- \nabla \cdot ( \bold{c} \otimes \nabla \bold{u}) + \bold{a} \bold{u} = \bold{f},
\end{equation*}

with user-supplied boundary conditions.

In our case, $\bold{a} = \bold{f} = \bold{0}$, and the coefficient tensor $\bold{c} = \begin{bmatrix} c_{11} & c_{12} \\ c_{21} & c_{22} \end{bmatrix}$ where
\[ c_{11} = \begin{bmatrix} \frac{E}{1-\nu^2} & 0 \\ 0 & \frac{E}{2(1+\nu)} \end{bmatrix},  c_{12} = \begin{bmatrix} 0 & \frac{\nu E}{1-\nu^2} \\ \frac{E}{2(1+\nu)} & 0 \end{bmatrix}, c_{21} = \begin{bmatrix} 0 & \frac{E}{2(1+\nu)} \\ \frac{\nu E}{1-\nu^2} & 0 \end{bmatrix}, c_{22} = \begin{bmatrix} \frac{E}{2(1+\nu)} & 0 \\ 0 & \frac{E}{1-\nu^2} \end{bmatrix}.\]

\section{Numerical computations}

In this appendix, we show how to approximate the integrals that appear in \eqref{loglike}. The averaged effective stress is computed in two-dimensional domain and assumed constant along the thickness dimension. Let $C \subset \mathbb{R}^2$ be the curve boundary of the specimen and $D_1 \subset \mathbb{R}^2$ be the projection of the specimen domain D into $\mathbb{R}^2$. Then, the integral $\int_{\partial D} F(n,\sigma^{\Delta}_{\mathrm{eff}}(\bold{x})) dS(\bold{x})$ is given by

\[ \int_{\partial D} F(n,\sigma^{\Delta}_{\mathrm{eff}}(\bold{x})) dS(\bold{x}) = \ell_{0} \int_{C} F(n,\sigma^{\Delta}_{\mathrm{eff}}(s)) ds + 2 \iint_{D_1} F(n,\sigma^{\Delta}_{\mathrm{eff}}(\bold{z}))\, d\bold{z}\,, \]
where $ \ell_{0}$ is the specimen thickness. We use the trapezoidal rule to approximate the line integral $\int_{C} F(n,\sigma(s)) ds$ where the discretization of $C$ is induced from the adaptive PDE solver. The second integral, $\int \int_{D_1} F(n,\sigma(\bold{z})) d\bold{z}$ is also approximated by trapezoidal rule using the FEM triangulation. In this case, the integral over the triangle $t$ with nodes $p_1, p_2$ and $p_3$ is simply 

\[ \frac{1}{3} \times Area(t) \times \left[ F(n,\sigma^{\Delta}_{\mathrm{eff}}(p_1)) + F(n,\sigma^{\Delta}_{\mathrm{eff}}(p_2)) + F(n,\sigma^{\Delta}_{\mathrm{eff}}(p_3)) \right] \, .\]


\bibliography{mybib}

\end{document}